# Unlocking tropical forest complexity: How tree assemblages in secondary forests boost biodiversity conservation

Maïri Souza Oliveira[a,b]*, Maxime Lenormand[a], Sandra Luque[a], Nelson A. Zamora[c], Samuel Alleaume[a], Adriana C. Aguilar Porras[d], Marvin U. Castillo[c], Eduardo Chacón-Madrigal[e], Diego Delgado[f], Luis Gustavo Hernández Sánchez[g], Marie-Ange Ngo Bieng[b], Ruperto Quesada-Monge[c], Gilberth S. Solano[h], Pedro M. Zúñiga[i]

[a]INRAE, National Research Institute on Agriculture, Food & the Environment, TETIS research unit, Maison de la télédétection, 34090 Montpellier, France

[b]CIRAD, Centre for International Cooperation in Agricultural Research for Development, Forests and Societies research unit, Montpellier 34398, France

[c]ITCR, Instituto Tecnológico de Costa Rica, Escuela de Ingeniería Forestal, Apartado 159-7050, Cartago, Costa Rica

[d]SINAC, Sistema Nacional de Áreas de Conservación, Departamento Conservación y Uso Sostenible de la Biodiversidad y los Servicios Ecosistémicos, MINAE, San Jose, Costa Rica

[e]Herbario Nacional, Museo Nacional de Costa Rica & Herbario Luis Fournier Origgi, Centro de Investigación en Biodiversidad y Ecología Tropical, Universidad de Costa Rica; San José, Costa Rica

[f]CATIE, Centro Agronómico Tropical de Investigación y Enseñanza, Turrialba 30501, Costa Rica

[g]UNA, Universidad Nacional, Instituto de Investigación y Servicios Forestales, Heredia, Costa Rica

[h]CODEFORSA, Comisión de Desarrollo Forestal de San Carlos, Costa Rica

[i]FUNDECOR, Fundación para el Desarrollo de la Cordillera Volcánica Central, Costa Rica

***Corresponding author:**

mairi.souza-oliveira@inrae.fr

Maison de la Télédétection, 500 rue Jean-François Breton, 34090 Montpellier, France

## Abstract

Secondary forests now dominate tropical landscapes and play a crucial role in achieving COP15 conservation objectives. This study develops a replicable national approach to identifying and characterising forest ecosystems, with a focus on the role of secondary forests. We hypothesised that dominant tree species in the forest canopy serve as reliable indicators for delineating forest ecosystems and untangling biodiversity complexity. Using national inventories, we identified in situ clusters through hierarchical clustering based on dominant species abundance dissimilarity, determined using the Importance Variable Index. These clusters were characterised by analysing species assemblages and their interactions. We then applied object-oriented Random Forest modelling, segmenting the national forest cover using NDVI to identify the forest ecosystems derived from in situ clusters. Freely available spectral (Sentinel-2) and environmental data were used in the model to delineate and characterise key forest ecosystems. We finished with an assessment of distribution of secondary and old-growth forests within ecosystems. In Costa Rica, 495 dominant tree species defined 10 in situ clusters, with 7 main clusters successfully modelled. The modelling (F1-score:

0.73, macro F1-score: 0.58) and species-based characterisation highlighted the main ecological trends of these ecosystems, which are distinguished by specific species dominance, topography, climate, and vegetation dynamics, aligning with local forest classifications. The analysis of secondary forest distribution provided an initial assessment of ecosystem vulnerability by evaluating their role in forest maintenance and dynamics. This approach also underscored the major challenge of in situ data acquisition.

## 1. Introduction

Tropical forests play a crucial role in preserving global biodiversity, harbouring a significant portion of terrestrial diversity, often estimated to be more than half of existing species(Myers et al. 2000). They are central to conservation strategies to achieve the Convention on Biological Diversity (COP15) goals, in particular the identification of priority ecosystems required to achieve the target of protecting 30% of land by 2030 (Mrema et al. 2020; CBD 2022; Pendrill et al. 2022). However, their degradation and deforestation remain among the leading drivers of global biodiversity loss (Bourgoin et al. 2024). As a result, many forest species are constrained to persist in human-modified landscapes, where forests survive within a matrix that varies significantly in its capacity to support biodiversity (Arroyo-Rodríguez et al. 2020). This raises critical questions about whether these anthropogenically influenced forested landscapes can sustain ambitious conservation objectives and at what spatial scale this support might be feasible (Perino et al. 2022).

In this context, secondary forests (SF), typically resulting from natural regeneration following anthropogenic pressures, primarily develop on fallow lands abandoned after agricultural use (Brown & Lugo 1990; Chazdon 2014). Today, SF constitute the majority of tropical forest cover, while old-growth forests (OGF) are increasingly restricted to inaccessible and non-arable areas (Edwards et al. 2019). While OGF, recognised for their complex structure and key role in sustaining biodiversity (Clark and Clark, 1996; Chazdon, 2014), SF are still often perceived as degraded systems (Pain et al., 2021). Although their species richness can recover relatively quickly, their species composition converges with that of OGF only over several centuries (Rozendaal et al. 2019; Poorter et al. 2021). Their regeneration dynamics are influenced by multiple interacting factors operating across spatial scales, contributing to the uncertainty of their trajectories (Walker et al. 2010; Arroyo-Rodríguez et al. 2017; Balvanera et al. 2021). As such, SF cannot replace OGF (Gibson et al. 2011), reinforcing the justification for prioritising OGF in conservation policies, often through "land-sparing" strategies (Mertz et al. 2021). However, this approach contributes to SF marginalisation, leaving them vulnerable to conversion into more economically lucrative land uses and undermining their potential contribution to long-term conservation efforts (Laurance et al. 2014; Reid et al. 2019). Located within human-modified tropical landscapes, SF are often cleared during the early stages of succession and struggle to fully restore ecosystem functions and biodiversity (Arroyo-Rodríguez et al., 2017). The limited recognition of the biodiversity value of these early stages and the uncertainty surrounding their trajectories, combined with conflicts between conservation and human land use, highlight a gap in conservation strategies regarding the integration of SF (Vieira et al. 2014; Sandoval et al. 2019; Shebitz et al. 2023). Nevertheless, despite the increase in protected forest areas, mainly OGF, contributing to biodiversity conservation, this measure remains insufficient to halt its rapid decline (Jenkins and Joppa, 2009; Gibson *et al.*, 2011; Barber *et al.*, 2014). Disturbances in the surrounding landscape continue to affect species within protected areas, calling into question the effectiveness of conservation strategies focused on OGF, particularly threatened by deforestation (Chazdon, 2014 ; Arroyo-Rodríguez *et al.*, 2020 ; Pain et al., 2021). In this context, SF are crucial as biological corridors, linking OGF fragments and facilitating genetic flow and species movement (Arroyo-Rodríguez et al. 2017). However, do SF have the capacity to preserve the specific flora composition that characterises tropical ecosystems, thereby complementing conservation efforts in OGFs?

To address this question and better understand the role of SFs in conserving tropical forest ecosystems and their contribution to international objectives, it is essential to develop robust and reproducible indicators and methods at the national scale capable of guiding the identification of priority ecosystems and supporting the implementation of conservation strategies aligned with global targets (Carroll & Noss 2022; Eckert et al. 2023). Nevertheless, despite numerous initiatives at international, continental, regional, and local levels, to develop such tools, their lack at the national scale remains a significant challenge (Perino et al. 2022; Shen et al. 2023).

Segmenting the forest cover into coherent ecosystems can guide conservation plans by integrating ecological and biological specificities unique to each forest ecosystem. Since the 1970s, studies on vegetation interactions with various environmental factors have shown that it reflects the complexity of biophysical processes (Monteith 1972; Droissart et al. 2018). Ecological, environmental, and historical filtering contribute to the distribution and spatial organisation of distinct species assemblages, representing large-scale biogeographical structures (Kreft & Jetz 2010; Araújo & Rozenfeld 2014), which influence the structure of ecosystems. Preserving the diversity of these ecosystems is essential to prevent the homogenisation of plant communities that structure them and combat biodiversity loss (Jakovac et al. 2022). This approach, applied to tropical forests, is documented in the scientific literature at different spatial scales, including regional (e.g., Moonlight et al. 2020), national (e.g., Pérez Chaves et al. 2020) and continental ( e.g., Jakovac *et al.*, 2022). However, it faces several limitations. The high complexity of elevated biodiversity and intricate species interactions in tropical forest ecosystems makes comprehensive biodiversity identification challenging (Pérez Chaves et al. 2020; Yang et al. 2023; Aguirre-Gutiérrez et al. 2025). In addressing this challenge, the trees that compose the forest canopy, as key structural elements of forest ecosystems, provide crucial insights into large-scale patterns of forest biodiversity (Rüger et al. 2020; Keppel et al. 2021; van Tiel et al. 2024). The specific composition of dominant tree species assemblages in forest canopy, in terms of abundance and structure, evolves based on forest succession stages, disturbance levels, and variations in environmental and historical factors (Chazdon 2008; Crouzeilles et al. 2016; Rosenfield et al. 2023). In tropical landscapes under anthropogenic pressures, the heterogeneity and turnover of these assemblages between plots are key factors determining forest resilience and stability (Lohbeck et al. 2016).

While ecological network analysis has mainly focused on species interactions at local scales (Schmid et al. 2020), methodological tools to assess interactions between ecosystems at national scales, particularly in tropical forests, remain underdeveloped. The approach proposed by Lenormand et al. (2019), applying species contributions and network theory to relationships between bioregions, provides an innovative avenue to address this gap by quantifying ecosystem interactions and specificity, thereby informing more integrated conservation strategies.

At the national scale, ecosystem delimitation is hindered by the lack of sufficient and reliable field data. The approach based on multispectral remote sensing, widely implemented through pixel-based classifications (e.g., Sentinel-2, Random Forest), has therefore become the dominant approach, with in situ data largely confined to calibration and validation roles (Massey et al. 2023; Saim & Aly 2025). However, the ecological value of the maps remains limited, constraining the information available for conservation (Pang et al. 2023). Multi-source approaches, combining in situ data with modelling based on spectral information, generate more interpretable classes that can reflect species assemblages, thereby enhancing their usefulness for management and conservation planning (Pérez Chaves et al. 2020; Pang et al. 2023). The integration of multisource data, including environmental variables, in this multi-source approach allows the limitations of spectral information alone—which under tropical conditions is often affected by cloud cover—to be overcome (Ferrer Velasco et al. 2022; Silveira et al. 2023). The use of object-oriented modelling based on spectral segmentation of the forest canopy enables the integration of data sources at different spatial resolutions and produces ecologically meaningful classes, while automating large-scale mapping (Waśniewski et al. 2020). This approach thus facilitates the operational delimitation of ecosystems at the national scale by incorporating ecological and environmental filters into precise and interpretable maps (Flores-Tolentino et al. 2021).

This study proposes a national-level approach for assessing the potential of SF in maintaining national tropical forest ecosystems. Our goal is to unravel the complexity of biodiversity by analysing the assemblages of tree species that shape and structure the forest canopy. The Costa Rica case study serves as the demonstrator for this approach. To achieve this, we identified and characterised high-level forest ecosystems and evaluated the distribution of OGF and SF across these ecosystems through four main steps: (i) using forest inventory data and botanical expertise, we identified clusters based on the dominant tree species of sites, (ii) we then characterised the clusters based on species distribution by identifying the associated assemblages of contributive species and analysing the interactions between these clusters, (iii) we modelled these clusters at the national forest cover level, using freely available spectral and environmental information to delineate and characterise key forest ecosystems defined by dominant species assemblages, and (iv) we assessed the distribution of forest-types (OGF and SF) within these ecosystems. We hypothesised (i) that the dominant tree species in the forest canopy serve as reliable indicators for delineating forest ecosystems at a national scale and (ii) that the interactions between forest ecosystems and the presence of SF are influenced by anthropogenic factors, linked to the specific environmental characteristics of each ecosystem. The developed pipeline for this multi-souce approach, is designed to be reproducible, innovates by relying on open multisource data (Sentinel-2, global environmental databases) and on well-established methods, but integrated in a novel way to enable the operational delineation of forest ecosystems at the national scale. To our knowledge, Pérez Chaves et al. (2020) also applied a multi-step approach to map floristic patterns at the national scale in Peru. However, the originality of the present study lies in its delineation and characterization of ecosystems based on canopy-dominant species assemblages, as well as its analysis of interactions among ecosystems. This approach is designed to translate ecosystem delineation into operational information for national conservation planning, grounded in the floristic patterns that structure the canopy, while incorporating ecological filters, local knowledge, and alignment with international conservation frameworks.

## 2. Material and Methods

### 2.1. A national-level demonstrator

Costa Rica, located in Central America, represents an exceptional world's biodiversity hotspot on just 0.03% of the Earth's surface, due to significant topographic and environmental gradients, as well as the ecological transition undertaken since the 1990s (Myers et al. 2000). The country is traversed by a central mountain range, composed of four main cordilleras (Guanacaste, Tilarán, Central, and Talamanca), creating a substantial altitudinal gradient, ranging from sea level to 3,820 meters above sea level and essentially two main slopes, Pacific and Caribbean (Figure 1b). Costa Rica also exhibits a broad environmental spectrum, ranging from the dry, highly seasonal biome of the northwestern Pacific coast to the very humid, seasonal biome of the southern Pacific coast and the nearly perennially wet conditions of the Caribbean slope. The major climatic gradient is along the Pacific slope, characterised by annual precipitation ranging from 1,475 mm in the northwestern to 5,070 mm in the southwestern (Figure 1c). Costa Rica has transitioned from a deforestation pioneer in the 1970s-1980s to a reforestation pioneer in the tropics by the 2000s (Redo et al. 2012). The current forest cover in Costa Rica reaches 52%, with 36% being SF (Figure 1a - Stan & Sanchez-Azofeifa 2019). However, this forest cover remains highly fragmented and mixed, with the landscape dominated by agricultural areas, pastures, and secondary forests (Stan & Sanchez-Azofeifa 2019).

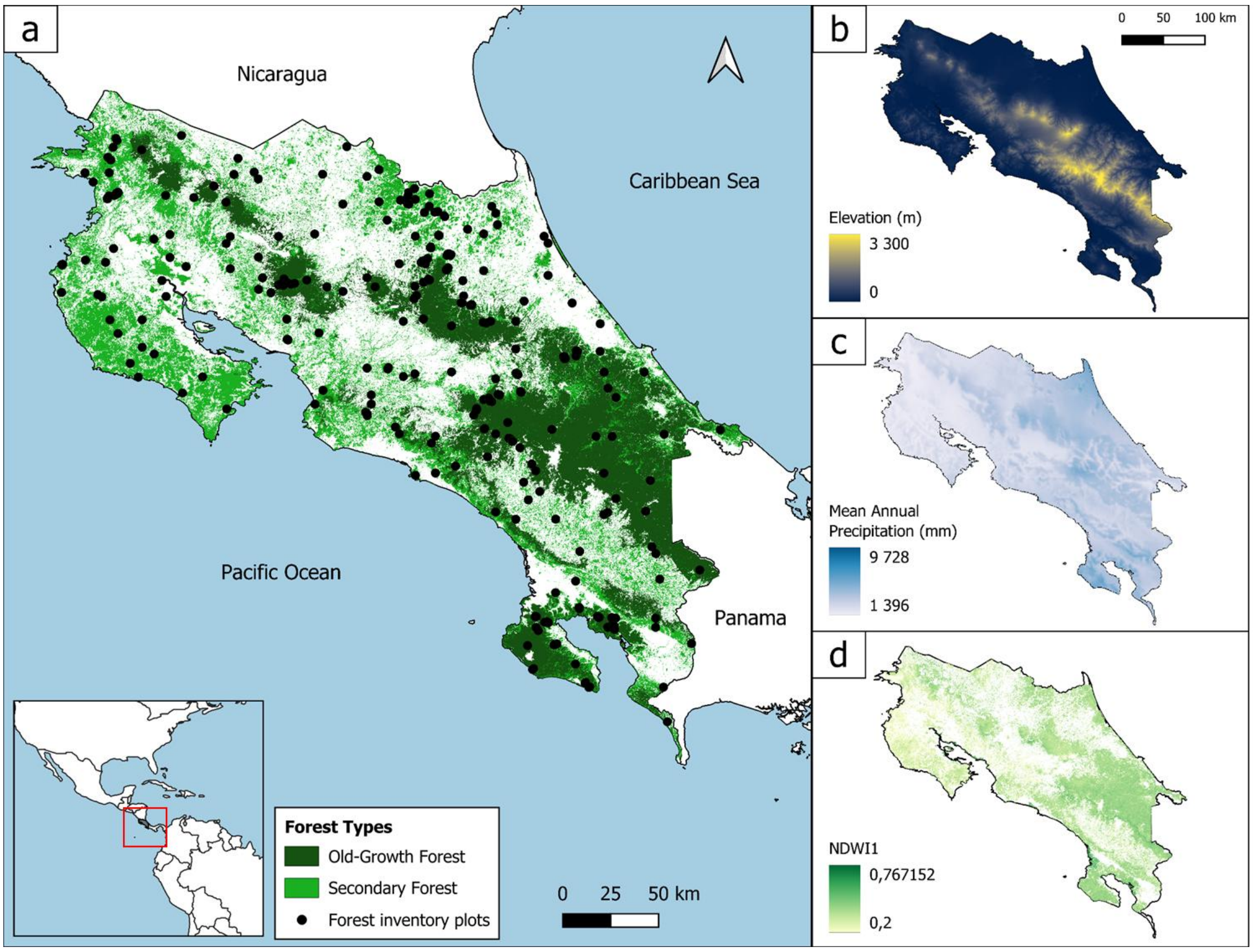

**Figure 1.** Presentation and environmental characteristics of the study area, Costa Rica: (a) Spatial distribution of 364 forest inventory plots across the national forest cover, (b) elevation (m), (c) Mean annual precipitation (mm/yr), (d) Normalised Difference Water Index (NDWI) calculated as the median of the time series over the most stable months of the 2018 dry season (February to April).

## 2.2. Data

**Tree species forest data –** We compiled data from the national inventory of Costa Rica (Programa REDD / CCAD-GIZ - SINAC, 2015) along with data from regional and local forest research projects. The resulting dataset includes 364 georeferenced plots (Figure 1a), ranging from 0.1 to 1.6 ha, carried out between 2004 and 2021. To standardise the data across the different inventories, we retained only trees with a DBH greater than 10 cm (DBH = diameter at breast height, at 1.3 m). Taxonomic identification was performed at the species level in all the selected plots. For the current study, we only accounted for individuals identified to the species level with confirmed taxonomic validation by botanical expertise. Consequently, our final database consists of 58,773 trees belonging to 1,333 species.

**Environmental data –** To define and characterise the identified forest ecosystems, we selected variables representing vegetation dynamic, topography, soil and climatic conditions. We represented vegetation dynamics during the dry and wet seasons by computing several vegetation spectral indices. These indices were computed, using multitemporal images from the Copernicus Sentinel-2 L2A satellite with 10m and 20m spatial resolution (CEOS 2024). These data were processed through Planetary Computer, which includes atmospheric correction using Sen2Cor (Main-Knorn et al. 2017). The vegetation indices were calculated by median over the multitemporal optical image series from 2018. To reduce spectral variability and ensure proper representation of the different seasons, two three-month periods were used: November to January for the wet season (Figure 1d) and February to April for the dry season. These vegetation indices are available as open access ("costarica-sentinel-2-l2-spectral-indices"). Elevation, along with several associated topographic metrics were derived from the NASADEM Merged DEM Global 1 Arc-Second V001, with a spatial resolution of 30m (NASA JPL 2020). Climate variables, with a spatial resolution of 1km, were obtained from CHELSA V1.0 (Karger et al. 2017), while soil variables, with a spatial resolution of 250m, were sourced from SoilGrids v0.5.5 (Hengl et al. 2017). The full set of variables used is presented in Table 1. We also used the 2021 forest-type map of Costa Rica, produced by SINAC at a spatial resolution of 10 m. This map was generated through a classification of Sentinel-2 and Sentinel-1 image mosaics (SINAC 2021) and is available as open access ("SNIT"). From this dataset, we selected SF and OGF forest-types (Figure 1a).

**Table 1.** Description of variables set used for this study

| Type | Variable | | Unit |
|---|---|---|---|
| Topography | DEM | Elevation | m |
| | Slope | — | degrees (°) |
| | Aspect | — | degrees (°) |
| | TWI | Topographic Wetness Index | — |
| | TRI | Terrain Ruggedness Index | m |
| | TPI | Topographic Position Index | — |
| Soil (0-30cm , 30-200cm, 0-200cm) | C | Soil Organic Carbon | dg/kg |
| | CEC | Cation Exchange Capacity | mmolc/kg |
| | Clay | — | g/kg |
| | N | — | cg/kg |
| | pH | — | — |
| | Silt | — | g/kg |
| Climate | anPR | mean annual precipitation | mm/year |
| | MAT | mean annual temperature | ◦C |
| | Prsea | precipitation seasonality (CV) | % |
| | Tsea | temperature seasonality (SD) | °C |
| Vegetation dynamic (Wet and Dry seasons) | CCI | Canopy Chlorophyll Content Index | — |
| | NDVI | Normalized Difference Vegetation Index | — |

NDWI Normalized Difference Water Index —

## 2.3. Biogeographical network analysis of in situ data

### 2.3.1. Identification of the dominant tree species in the forest canopy

Our objective was to identify the global floristic patterns of the forest canopy across the country. Due to the complexity of tropical forest diversity and the local biodiversity richness, some species were expected to have low prevalence, which could bias the delineation of forest ecosystems. To avoid this issue, we used the Importance Value Index (IVI) to determine which species contributed most to each plot's composition, structure, and dynamics of the tree communities (Figure 2a). The IVI combines three parameters: (i) relative density, measured by abundance, (ii) relative dominance, measured by basal area, and (iii) relative frequency, which represents the proportion of subplots where the species are present (Curtis & McIntosh 1951).

We set an IVI threshold of 5% to select the dominant species in each forest plot, which was then used for the forest ecosystems delineation. We performed a sensitivity analysis to evaluate the impact of selecting a 5% IVI threshold on the delineation of forest ecosystems. For each plot, we calculated the cumulative IVI captured by species exceeding thresholds of 5% and 10%, and found that the 5% threshold retained on average 60% of the cumulative IVI, whereas the 10% threshold retained only 47% (Figure S1).We also assessed the stability of the forest cluster composition using pairwise comparisons (Adjusted Rand Index, ARI) between clustering results obtained with thresholds of 0%, 5%, and 10%. The ARI values showed high similarity between the 5% and 10% thresholds (ARI = 0.64), indicating that the overall cluster structure is robust to variations in the threshold. These results support the choice of a 5% threshold for selecting dominant species, as it captures the majority of canopy dominance, balances the inclusion of the most influential species, ensures stability of the clustering results, and avoids excluding moderately abundant species.

### 2.3.2. Hierarchical clustering of sites

In this study, we considered forest inventory plots as sites. The delineation of forest ecosystems was based on automatic discrimination of tree species assemblages, aiming to minimise the taxonomic turnover rate within ecosystems while maximising it between them (Kreft & Jetz 2010). We performed hierarchical clustering based on the dominant tree species assemblages to cluster the sites according to their floristic similarities in an unsupervised approach (i.e., without incorporating environmental variables) (Figure 2a). To achieve this, we applied the UPGMA (Unweighted Pair Group Method with Arithmetic Mean) agglomerative algorithm using the turnover component of the Bray-Curtis dissimilarity index ($\beta\ \mathrm{BrayTurn}$). $\beta\ \mathrm{BrayTurn}$ specifically focuses on changes in species composition by quantifying the minimal difference in presence and abundance. This approach emphasises the species unique to each site (Eq. 1) (Baselga 2013). This metric is particularly suitable for our data because it allows the analysis of variations in species composition without being influenced by abundance differences related to site size.

$$\beta\ \mathrm{BrayTurn} = 1 - \frac{\min(B,C)}{A+\min(B,C)} \qquad \text{(Eq. 1)}$$

In Eq. 1, $A$ represents the sum of the minimum abundances of species shared between two sites, while $B$ and $C$ correspond to the abundances of species present exclusively on each site of the pair. The function $\min(B, C)$ measures the turnover between the two sites, i.e., the total number of individuals of species present only on one of the two sites.

The UPGMA method has proven more effective for detecting consistent biogeographical patterns than other hierarchical classification methods and does not weight clusters based on their size (Kreft & Jetz 2010). Since this method is influenced by site order in the distance matrix, we performed 1000 randomisations of the order in the dissimilarity matrix (Dapporto et al. 2013). To quantitatively assess each resulting tree, we calculated the cophenetic correlation coefficient, which measures the correlation between the distance at which the sites are connected in the tree and the distance between the sites in the initial dissimilarity matrix (Sokal & Rohlf 1962).

The optimal number of clusters was selected using the Elbow Method based on the pc_distance metric, in order to maximise the dissimilarity between clusters. This metric calculates the ratio between the sum of β dissimilarities between clusters and the total sum of β dissimilarities for the entire dissimilarity matrix (Holt et al. 2013).

Single-site clusters were further assessed for their representativeness using pairwise similarity tests based on species composition (Simpson's distance with Bonferroni correction) between the isolated sites and the main clusters. A 50% similarity threshold, supported by botanical expertise, was used to determine whether isolated sites represented genuine floristic specificity or simply exhibited a high rate of unique species due to random sampling and could therefore be associated with existing clusters (see Table S1). Isolated plots deemed not representative of broader floristic patterns were excluded from further analysis, as we did not have a sufficient number of plots to adequately represent these specific floras in the modelling.

### 2.3.3. Test value matrix

To analyse the contribution of species to each cluster and characterise the associated species assemblages, we used test values ρ, which measure the under- or over-representation of species in the clusters (Lebart et al. 2000) (Figure 2b). The test value ρ for a species i in a cluster j quantifies the difference between the mean abundance $\mu_{ij}$ of species $i$ in $n_j$ samples from cluster j and its mean abundance $\mu$ in $n$ samples from the entire study area, standardised to assess whether the species is over- or under-represented in this cluster compared to all samples (Eq. 2). Since this quantity depends on the size of the clusters ($n_j$), it is normalised by the standard deviation associated with the expected mean abundance if the variability in cluster j were comparable to that of the total population represented by {n, μ, σ²}, taking into account the difference in cluster sizes.

$$\rho_{ij} = \frac{\mu_{ij} - \mu}{\sqrt{\frac{n-n_j}{n-1} \times \frac{\sigma^2(X)}{n_j}}} \qquad \text{(Eq. 2)}$$

To define the representative species in the assemblages associated with the clusters, we qualified the species that contributed positively and significantly to one or more clusters by introducing a significance threshold δ applied to the test values ρ corresponding to a one-tailed significance level of 2.5% in a Gaussian distribution, i.e., $\delta$=1.96. Thus, the matrix of test values $\rho$ highlights the sets of species that best characterise the clusters (Lenormand et al. 2019).

**2.3.4. Interaction network between clusters**

To quantify the relationships between the clusters, we examined the distribution among clusters of species with a contributions were significantly positive ($\rho_{ij} \geq 1.96$)(Lenormand et al. 2019). $\rho^+$ that represents these significantly positive contributions of species to the clusters (Figure 2b). To obtain the relative contribution $\hat{\rho}^+_{ij}$ of a species $i$ to a cluster j, we normalised the contributions $\rho^+$ by the sum of $\rho^+$ for all clusters $k$ (Eq. 3).

$$\hat{\rho}^+_{ij} = \frac{\rho^+_{ij}}{\sum_k \rho^+_{ik}} \qquad \text{(Eq. 3)}$$

Subsequently, we calculated for each cluster j the mean fraction of contribution to the cluster from species that also contribute significantly to cluster j′, denoted as $\lambda_{ij'}$, based on Eq. 4. In this equation, $A_j$ represents the set of species for which the contribution to cluster j is significant, i.e., those for which $\rho_{ij} \geq 1.96$. The normalisation of contributions by $|A_j|$ ensures that the similarity measure is independent of the size of $A_j$. Thus, $\lambda_{jj}$ expresses the specificity of cluster j. These fractions are expressed as percentages with a vector $\lambda_j$ for a given cluster that sums to 1. Therefore, the values of $\lambda_j$ provide measures of both specificity and connectivity between the clusters, based on the representative species they share.

$$\lambda_{jj'} = \frac{1}{|A_j|} \sum_{i \in A_j} \hat{\rho}^+_{ij'} \qquad \text{(Eq. 4)}$$

We then used the values from the matrix to construct an interaction network illustrating the relationships between site clusters and the associated species assemblages. This network was then projected into a two-dimensional space, using altitude derived from the DEM and annual precipitation, taking the median of these variables for each cluster, in order to explore the main environmental gradients of the country.

## 2.4. Delimitation of forest ecosystems using spectral and environmental data

**2.4.1. Data standardisation**

To address the difference in spatial resolution between floristic and environmental data, we used segmentation to create an object-oriented approach (Figure 2c). This process divided the forest cover, derived from the national forest-type map, into spectrally uniform segments, helping to standardise the data for ecosystem modelling and validating environmental factors. The segmentation was based on an NDVI mosaic derived from the median of Sentinel-2 images from the dry season of 2018, using the Large Scale Generic Region Merging (LSGRM) algorithm from the Moringa Land Cover Toolbox (Gaetano et al. 2019). Subsequently, for each segment, we performed statistical zoning by extracting the 25$^{th}$ (q25), 50$^{th}$ (q50), and 75$^{th}$ (q75) percentiles for each numerical explanatory variable. This approach reduced biases from local variability and captured environmental trends representative of each segment, aiding in the characterisation of forest ecosystems. Each segment containing a site adopted the cluster value and species assemblage of that site, based on the assumption that it accurately represents the entire segment.

**2.4.2. Modelling clusters at the national scale**

A Random Forest model (Breiman 2001) was developed using the clustering results from the sites to predict the clusters within the Costa Rica forest cover, aiming to produce a map of the main forest ecosystems (Figure 2c). For the modelling, we selected only the main clusters with a minimum of 15 sites. Most of the 34 predictive variables were highly correlated. To limit overfitting, we applied a two-step variable selection procedure. First, we used a recursive feature elimination (RFE) procedure with a Random Forest model, optimising the F1-score to identify the best-performing subsets of predictors (sizes between 5 and 30 variables were tested). The best-performing subset was then examined for multicollinearity using Pearson's correlation coefficients. Variables with $|r| > 0.7$ were considered redundant and manually removed based on ecological relevance and interpretability. Model evaluation was performed using 10-fold cross-validation, with performance assessed through the F1-score and the macro F1-score. The macro F1-score, calculated as the arithmetic mean of the F1 scores for all classes, which treats all classes equally, is mainly well suited to unbalanced classes (Sokolova & Lapalme 2009). For each predicted forest ecosystem, we assessed the variable importance using 100 permutation per variable (Ramosaj & Pauly 2023), as well as the marginal effects of the most important variables by constructing partial dependence plots (Greenwell 2017). This final step allows to better characterise these ecosystems and establish links with botanical expertise.

## 2.5. Analysis of the distribution of forest-types within predicted forest ecosystems

We calculated the proportions of SF and OGF in the *in situ* clusters and the derived forest ecosystems to assess the concordance between the prediction results and clustering results and to analyse their distribution across different forest ecosystems (Figure 2d). To do this, the vector layer of the main Costarican forest ecosystems was converted into a raster with a spatial resolution of 10 m to be compatible with the forest cover raster of the same resolution. The proportions of forest-types were then calculated based on the number of pixels corresponding.

All analyses were performed using R version 4.3.3 (R Core Team 2024), with several specific packages, including "bioregion" (Denelle et al. 2025) for biogeographical network analysis, "caret" (Kuhn et al. 2024) for variable selection, training and prediction of the Random Forest model, and "rfPermute" (Archer 2023) for variable importance assessment, and finally the "pdp" package (Greenwell 2024) for calculating the marginal effects of the variables.

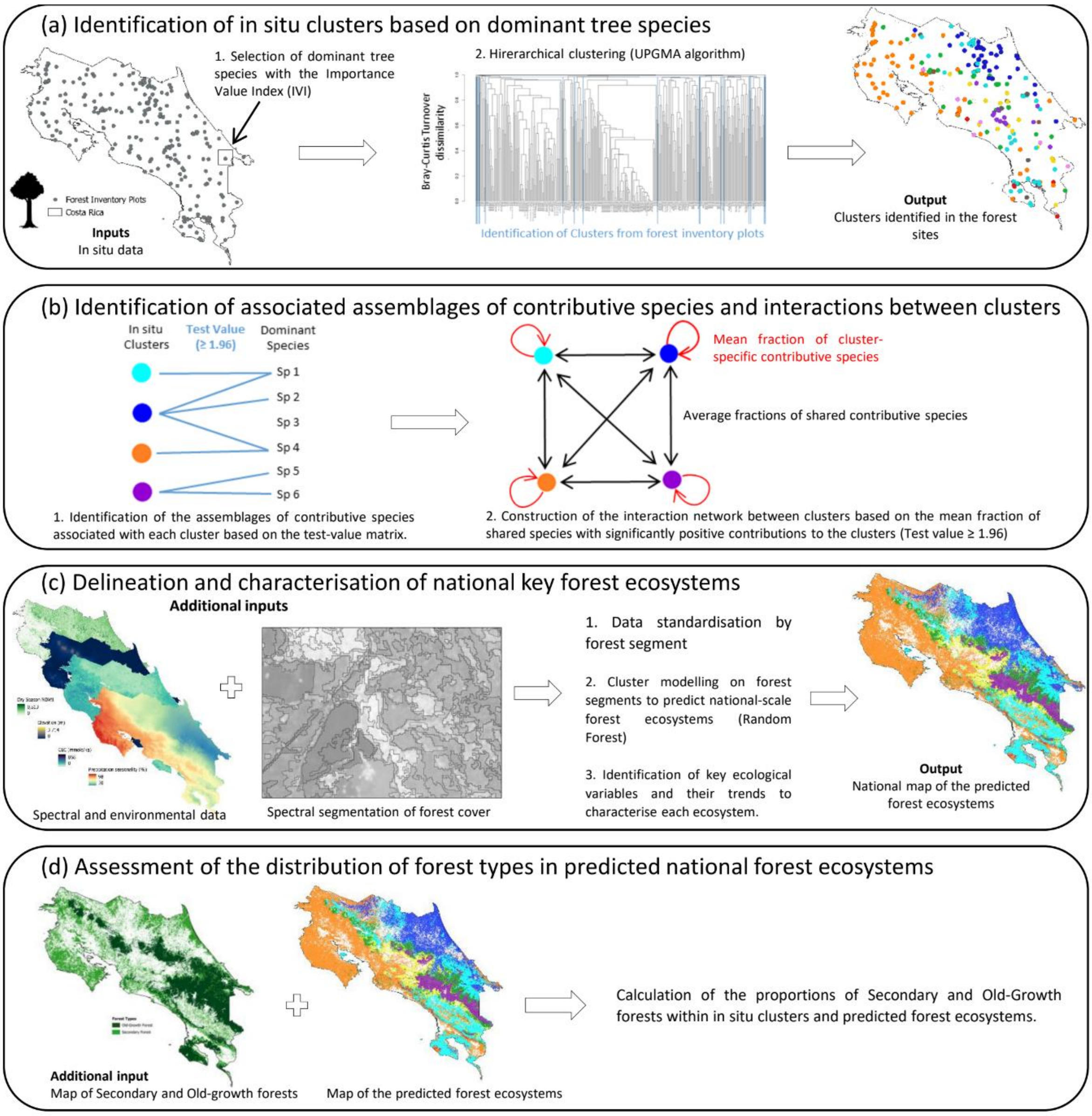

**Figure 2.** Schematic workflow used in this study, illustrating the key methodological steps.

## 3. Results

### 3.1. Biogeographical network analysis

We identified 18 optimal clusters from the hierarchical clustering analysis applied to the forest inventory data. This analysis was based on the distribution of 495 dominant tree species selected from the 1333 species present at the sites, using a threshold IVI of 5%. These clusters explained 85.74% of the observed variation in species composition (Figure S2a). However, 8 of these clusters were not usable, represented by a single site, and were either merged with the main clusters if they showed sufficient floristic similarity, or excluded when they did not meet the representativeness threshold (details in Table S1). As a result, we retained 10 clusters for subsequent analysis (Figure 3a) that explained 69.77% of the observed variation (Figure S2). The cluster sizes ranged from 4 to 101 sites. These clusters, although highly unbalanced (Figure 4b), were obtained using the UPGMA algorithm, which preserves the structure of the dissimilarity matrix.

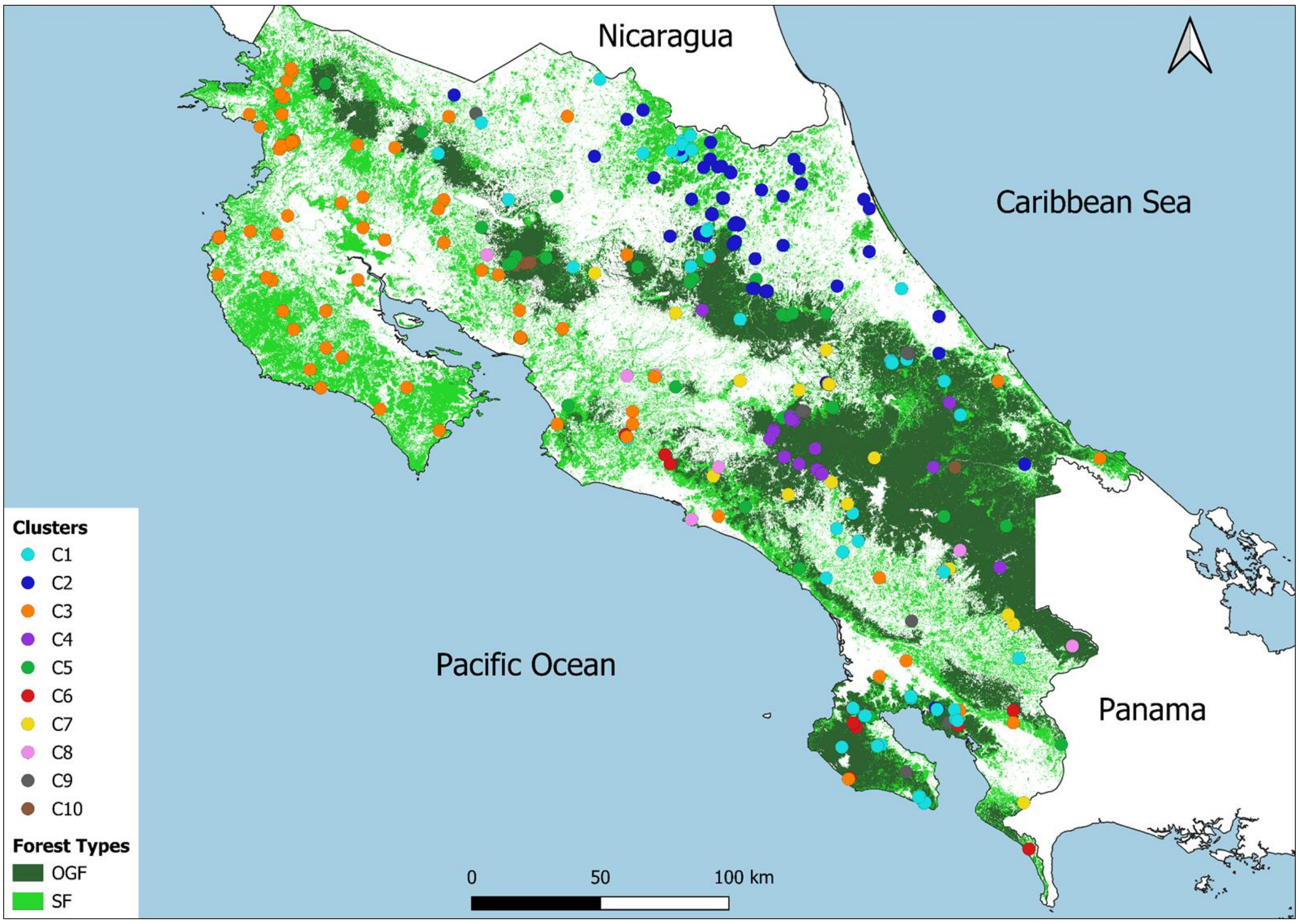

**Figure 3.** Geographical distribution of the 10 identified site clusters, determined through hierarchical clustering based on dissimilarity in species composition.

According to the significance threshold δ set at 1.96, 71% of the species contributed to a single cluster, and 18% contributed to two clusters. The maximum number of species contributing to only one cluster was reached at the specified threshold, indicating that the majority of the dominant canopy species were mainly associated with a single cluster (Figure 4a). As a result, the species assemblages associated with the clusters included between 20 and 88 contributive species, i.e., species that contributed positively and significantly to the clusters (Figure 4b). Details of the contributive species and their contribution values $\rho_{ij}$ are available in Table S2. In total, 322 species were identified as contributive among the 495 species dominating the forest canopy. The interaction network between clusters, based on the fraction of contributing species ($\lambda_{jj'}$), validated the strong specificity of the clusters, with a mean of 79.2%, ranging from 63% for cluster C9 to 98% for cluster C3 (Figure 5). Despite these variations in specificity, the network revealed weak interactions between clusters, with the maximum sharing of contributing species reaching 13% for $\lambda_{6,1}$. Clusters C5, C6, C9, and C10 appear as transition zones between clusters associated with the high, medium, and low-altitude wet biomes. In contrast, cluster C3, isolated, seems representative of the low-altitude dry biome. All $\lambda_{jj'}$ interactions and shared species are presented respectively in Tables S3 and S4.

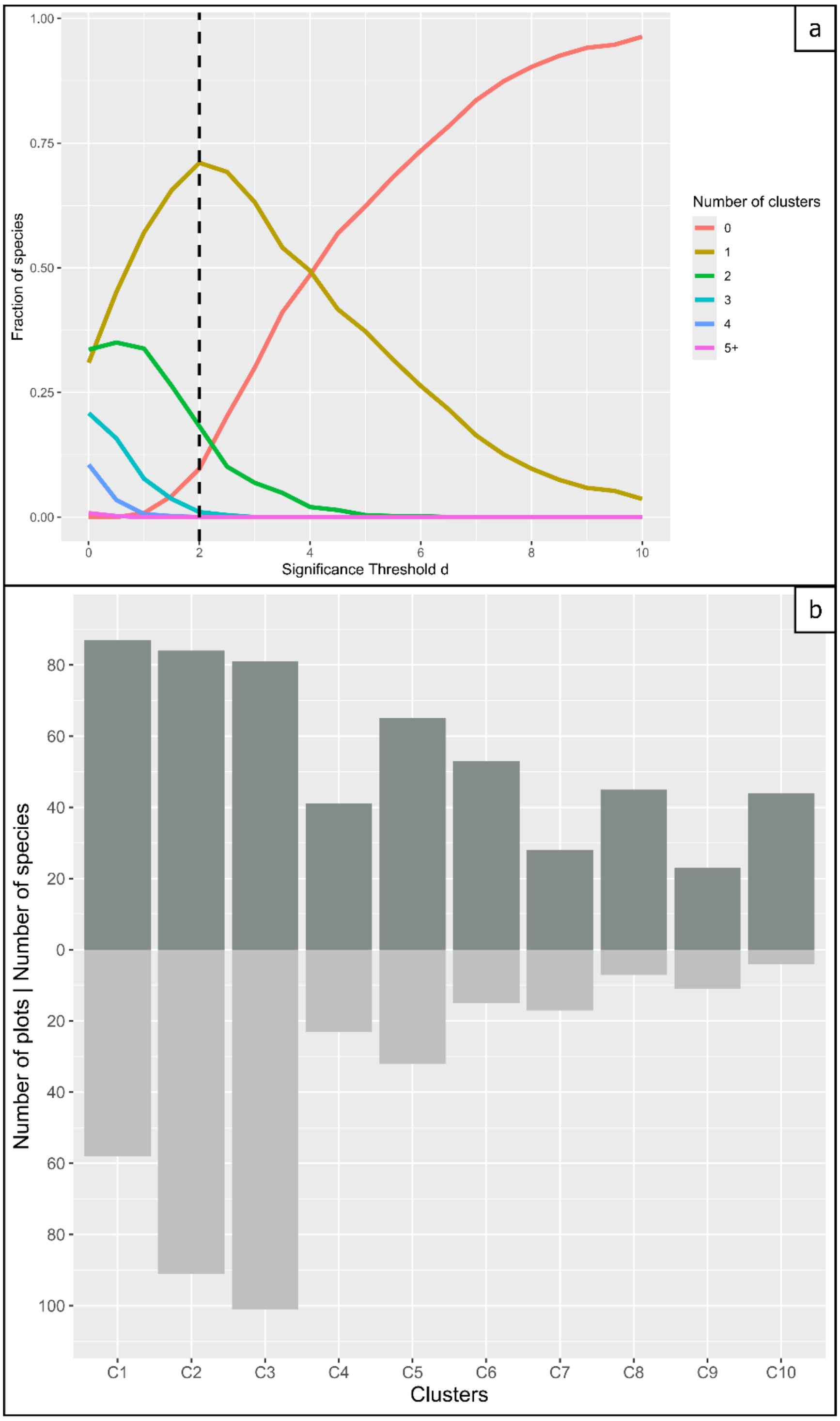

**Figure 4.** Analysis of species contribution to identified clusters: (a) Fraction of species contributing positively and significantly to a given number of clusters (from 0 to 5 or more) as a function of the significance threshold. The vertical line represents the significance threshold δ = 1.96. (b) Barplot of the numbers of contributive species ($\rho_{ij}$≥1.96) and sites according to identified clusters.

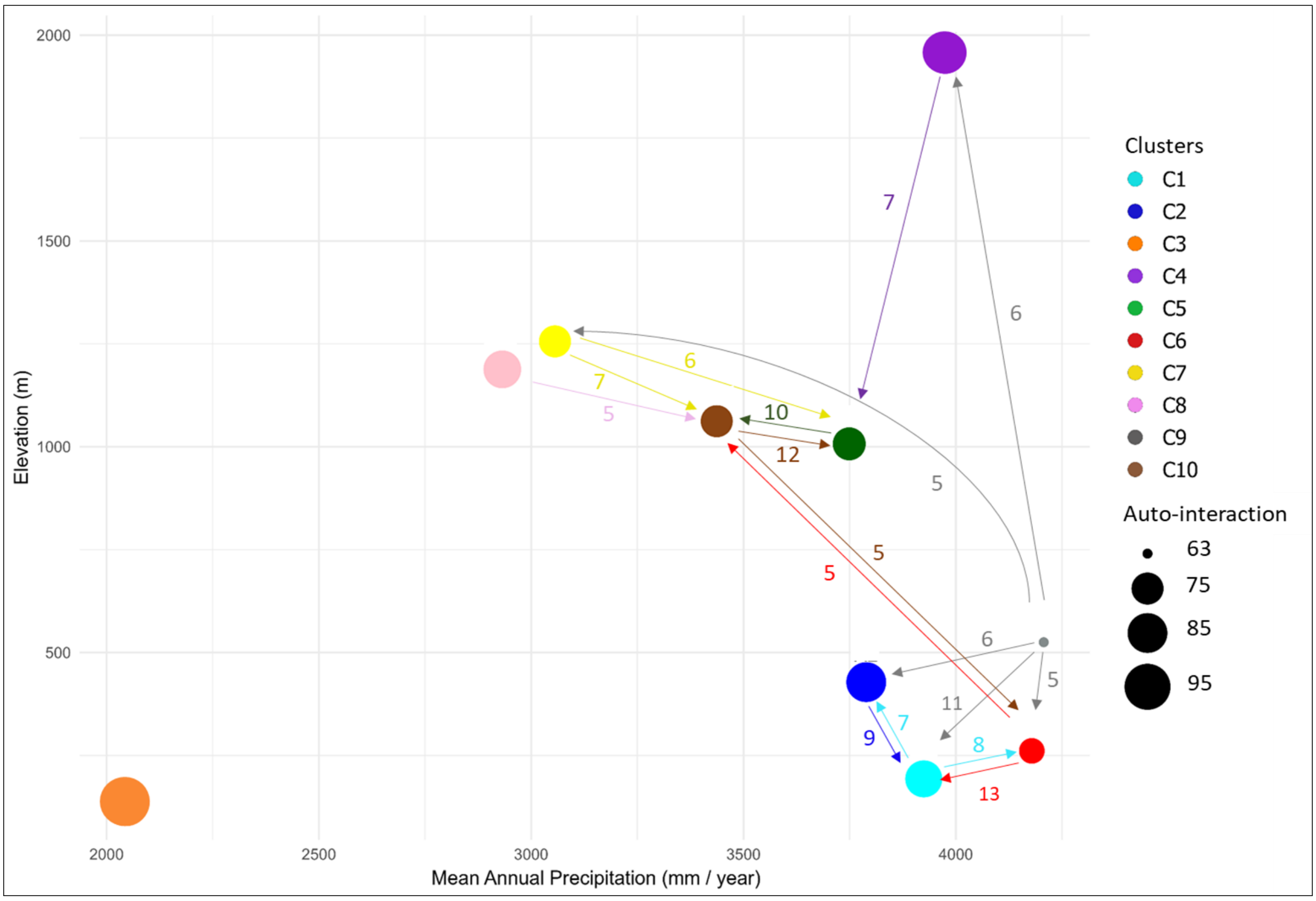

**Figure 5**. Interaction network illustrating the relationships between clusters and the associated contributive species assemblages. $\lambda_{jj'}$, expressed as a percentage, represents the mean fraction of contribution to cluster j of species that also contribute significantly to cluster j′. Only interaction with a value $\lambda_{jj'}$ higher than 5% are shown. The network is represented in two-dimensional space, with altitude (in m) and Mean Annual Precipitation (in mm/year). The median of these variables is used for each cluster.

### 3.2. Delimitation and characterisation of forest ecosystems using spectral and environmental data

Only clusters C1 to C7 had a sufficient number of plots to be modelled across the forest cover. The random forest classification models showed optimal performance when they included only the variables selected through a recursive feature elimination (RFE) procedure combined with manual removal of collinearity (Figure S3). The model achieved an overall F1-score of 0.73 and macro F1-score of 0.58, although performance varied significantly across the clusters. Significant classification errors were observed for clusters C5, C7, and particularly C6, with respective error rates of 52.7%, 66.7%, and 83.3% (Table 2). Frequent classification errors observed in the confusion matrix, particularly from C1 to C2, from C6 to C1, from C4 to C5, as well as from C7 to C5 and C4, align with the cluster interactions highlighted in previous analyses. A notable proportion of classification errors was also observed from C5 to C1. In contrast, clusters C2 and C3 were distinguished with high accuracy by the model.

**Table 2.** Assessment of random forest model performance: confusion matrix expressed in percentages and F1-Score by class for the best model with a global F1-Score of 0.73 and a macro F1-score of 0.58.

| | C1 | C2 | C3 | C4 | C5 | C6 | C7 | **F1-Score** |
|---|---|---|---|---|---|---|---|---|
| C1 | 63.74 | 10.53 | 7.60 | 0.00 | 5.85 | 7.02 | 5.26 | 0.60 |
| C2 | 7.04 | 86.38 | 5.63 | 0.00 | 0.94 | 0.00 | 0.00 | 0.83 |
| C3 | 7.50 | 6.67 | 81.67 | 0.00 | 0.83 | 2.50 | 0.83 | 0.81 |
| C4 | 1.59 | 0.00 | 0.00 | 69.84 | 22.22 | 0.00 | 6.35 | 0.74 |
| C5 | 17.20 | 9.68 | 9.68 | 6.45 | 47.31 | 6.45 | 3.23 | 0.50 |
| C6 | 58.33 | 0.00 | 12.50 | 0.00 | 12.50 | 16.67 | 0.00 | 0.19 |
| C7 | 9.52 | 9.52 | 11.90 | 14.29 | 14.29 | 7.14 | 33.33 | 0.38 |

This classification model allowed to define, characterise, and delineate the forest ecosystems of Costa Rica based on the seven main clusters modelled. The seven explanatory variables selected for the model input show varying levels of discrimination depending on the clusters (Table 3). According to Mean Decrease Accuracy, which measures the impact of a variable on model accuracy by randomly permuting its values, all variables are globally significant. Altitude (q50.DEM) is the most discriminant variable on average, followed by precipitation seasonality (q25.PRSea), the NDWI index during the wet season (q25.NDWIw), and annual precipitation (q75.anPR). The least discriminant variables on average are pH (q50.pH30), CEC (q50.CEC30) of the topsoil layer, and slope (q25.slope). According to Mean Decrease Gini, which measures the reduction in node impurity in decision trees when a variable is used for splitting, only four variables are significantly discriminant, precipitation seasonality (q25.PRSea) and altitude (q50.DEM), followed by annual precipitation (q75.anPR) and soil pH (q50.pH30). The variable importance for each cluster highlights the model's difficulty in predicting clusters C5 to C7 using topographic, environmental, and seasonal vegetation dynamics variables.

**Table 3.** Variable importance in the random forest of the seven variables selected as model inputs with (i) the variable importance value from the permutation test with its statistical significance (p-value) for each cluster and (ii) the Mean decrease accuracy and Mean decrease gini metrics with their statistical significance (p-value) for the overall model. Stars (*) indicate statistically significant importance values ($p < 0.05$); shading reflects the relative importance (Imp) of each variable, with darker colours representing higher importance.

| **Variables** | **C1** | | **C2** | | **C3** | | **C4** | | **C5** | | **C6** | | **C7** | | **Mean Decrease Accuracy** | | **Mean Decrease Gini** | |
|---|---|---|---|---|---|---|---|---|---|---|---|---|---|---|---|---|---|---|
| | **Imp** | **p_value** | **Imp** | **p_value** | **Imp** | **p_value** | **Imp** | **p_value** | **Imp** | **p_value** | **Imp** | **p_value** | **Imp** | **p_value** | **Value** | **p_value** | **Value** | **p_value** |
| q50.DEM | 15.10 | 0.01 * | 25.66 | 0.01 * | 11.77 | 0.01 * | 35.50 | 0.01 * | 17.08 | 0.01 * | 2.76 | 0.14 | 13.65 | 0.01 * | 41.87 | 0.01 * | 38.16 | 0.01 * |
| q25.PRSea | 15.42 | 0.01 * | 35.39 | 0.01 * | 20.23 | 0.01 * | 14.42 | 0.01 * | 4.61 | 0.04 * | 1.19 | 0.26 | 6.08 | 0.02 * | 37.66 | 0.01 * | 40.64 | 0.01 * |
| q25.NDWIw | 18.37 | 0.01 * | 24.36 | 0.01 * | 7.42 | 0.07 | 1.22 | 0.31 | 10.53 | 0.01 * | 5.55 | 0.02 * | 8.97 | 0.01 * | 30.25 | 0.01 * | 30.38 | 1.00 |
| q75.anPR | 17.36 | 0.01 * | 13.18 | 0.01 * | 21.07 | 0.01 * | 11.48 | 0.01 * | 5.49 | 0.02 * | 1.70 | 0.20 | 4.93 | 0.06 | 27.24 | 0.01 * | 37.20 | 0.01 * |
| q50.pH30 | 15.19 | 0.01 * | 15.37 | 0.01 * | 18.45 | 0.01 * | 15.48 | 0.01 * | -3.43 | 0.79 | 0.06 | 0.39 | 8.04 | 0.01 * | 25.37 | 0.01 * | 34.79 | 0.05 * |
| q50.CEC30 | 6.71 | 0.03 * | 18.79 | 0.01 * | 11.76 | 0.01 * | 4.43 | 0.05 * | 6.99 | 0.03 * | 0.28 | 0.38 | 4.77 | 0.04 * | 22.45 | 0.01 * | 26.93 | 1.00 |
| q25.slope | 5.46 | 0.07 | 20.89 | 0.01 * | -3.57 | 0.91 | 11.83 | 0.01 * | -2.60 | 0.79 | -0.59 | 0.46 | 7.08 | 0.01 * | 20.03 | 0.01 * | 24.10 | 1.00 |

Table 4 presents the identification and characterisation of the predicted forest ecosystems by combining (i) the significant variable importance (Table 3) and the marginal effects of key ecological variables retained by the model, whose log-odds define the main ecological trends (Figure S4), (ii) the analysis of contributive species associated with the clusters (Table S2), (iii) the mean values of each explanatory variable for each cluster (Table S5), and (iv) the distribution of the seven forest ecosystems across Costa Rica (Figure 6a). Although the model's performance is relatively weak for clusters C5 to C7, it still captured certain ecological gradients that contributed to the definition of the associated forest ecosystems: PMC-C, TWPE-P, and PMC-P.

**Table 4.** Correspondence between clusters, modelled forest ecosystem names, and associated acronyms, with characterisation based on key ecological variables and contributive species assemblages: Only ecological variables retained by the model with a significant importance score (p_value ≤ 0.05) were considered for characterisation of each ecosystem.

| Clusters | Ecosystems | Acronyms | Characterisation<br>Contributive species | Tropography | Climate | Vegetation dynamic | Soil | Location |
|---|---|---|---|---|---|---|---|---|
| C1 | Wet Seasonal Evergreen forest | WSE | Majority of evergreen species (predominated by *Elaeoluma glabrescens, Symphonia globulifera* and *Garcinia madruno*) | Broad median elevation range from low to intermediate (504 ± 361 m) | Maximum mean annual precipitation of 3,611 ± 707 mm/yr<br><br>Minimum precipitation seasonality variation of 51 ± 13%. | Minimum NDWI range during the wet season of 0.31 ± 0.11 | Median pH of 5.21 ± 0.62<br><br>Median CEC of 134 ± 44 mmolc/kg | Southern Pacific and Caribbean coasts |
| C2 | Lowland Wet Evergreen forest of Caribbean slope | LWE-C | Evergreen species (predominated by *Pentaclethra macroloba*) | Low median elevation (≤346 m)<br><br>Low minimal slope (4 ± 4°) | Maximum mean annual precipitation of 3,522± 726 mm/yr<br><br>Minimum precipitation seasonality variation of 34 ± 9% | Minimum NDWI range during the wet season of 0.31 ± 0.07 | Median pH of 5.27 ± 0.62<br><br>Median CEC of 132 ± 49 mmolc/kg | Caribbean coast |
| C3 | Lowland Dry-to-Moist Deciduous-to-Semi-deciduous forest | LDM-DS | Mainly composed of deciduous to semi-deciduous species (predominated by *Handroanthus ochraceus, Bursera simaruba* and *Spondias mombin*) | Low median elevation (≤519 m) | Maximum mean annual precipitation of 2,308 ± 542 mm/yr<br><br>Minimum precipitation seasonality variation of 75 ± 14% | Minimum NDWI range during the wet season of 0.22 ± 0.09 | Median pH of 5.73 ± 0.69<br><br>Median CEC of 212 ± 60 mmolc/kg | Mainly on the southern Pacific coast |
| C4 | Mountain Oak Rainforest | MOR | Predominantly composed of oak species (dominated by *Quercus sapotifolia*) | High median elevation (2,269 ± 463 m)<br><br>High minimal slope (20 ± 8°) | Maximum mean annual precipitation of 3,804 ± 614 mm/yr)<br><br>Minimum precipitation seasonality variation of 45 ± 10% | — | Median pH of 5.12 ± 0.17<br><br>Median CEC of 192 ± 39 mmolc/kg | Talamanca cordillera |
| C5 | Premontane-to-mountain Mixed-to-evergreen Cloud forest of Caribbean slope | PMC-C | Mix of evergreen and deciduous species with a large altitudinal gradient (predominated by *Ruagea glabra*, *Elaeagia auriculata*, *Salacia petenensis*, *Cecropia angustifolia* and *Inga oerstediana*) | Intermediate median elevation (1,142 ± 361 m) | Maximum mean annual precipitationof 3,616 ± 547 mm/yr<br><br>Minimum precipitation seasonality variation of 48 ± 13% | Minimum NDWI range during the wet season of 0.37 ± 0.05 | Median CEC of 171 ± 45 mmolc/kg | Caribbean slope |
| C6 | Transitional wet premontane evergreen forest of the Pacific slope | TWPE-P | Evergreen species (predominated by *Otoba novogranatensis*) | — | — | Minimum NDWI range during the wet season of 0.39 ± 0.03 | — | Southern Pacific coast |
| C7 | Premontane-to-mountain Mixed-to-evergreen Cloud forest of Pacific slope | PMC-P | Mix of evergreen and deciduous species with a large altitudinal gradient (predominated by *Saurauia montana* and *Inga punctata*) | Intermediate median elevation (1,360 ± 372)<br><br>High minimal slope (16 ± 8°) | Minimum precipitation seasonality variation of 55 ± 12% | Minimum NDWI range during the wet season of 0.26 ± 0.08 | Median pH of 5.29 ± 0.31<br><br>Median CEC of 162 ± 33 | Pacific slope |

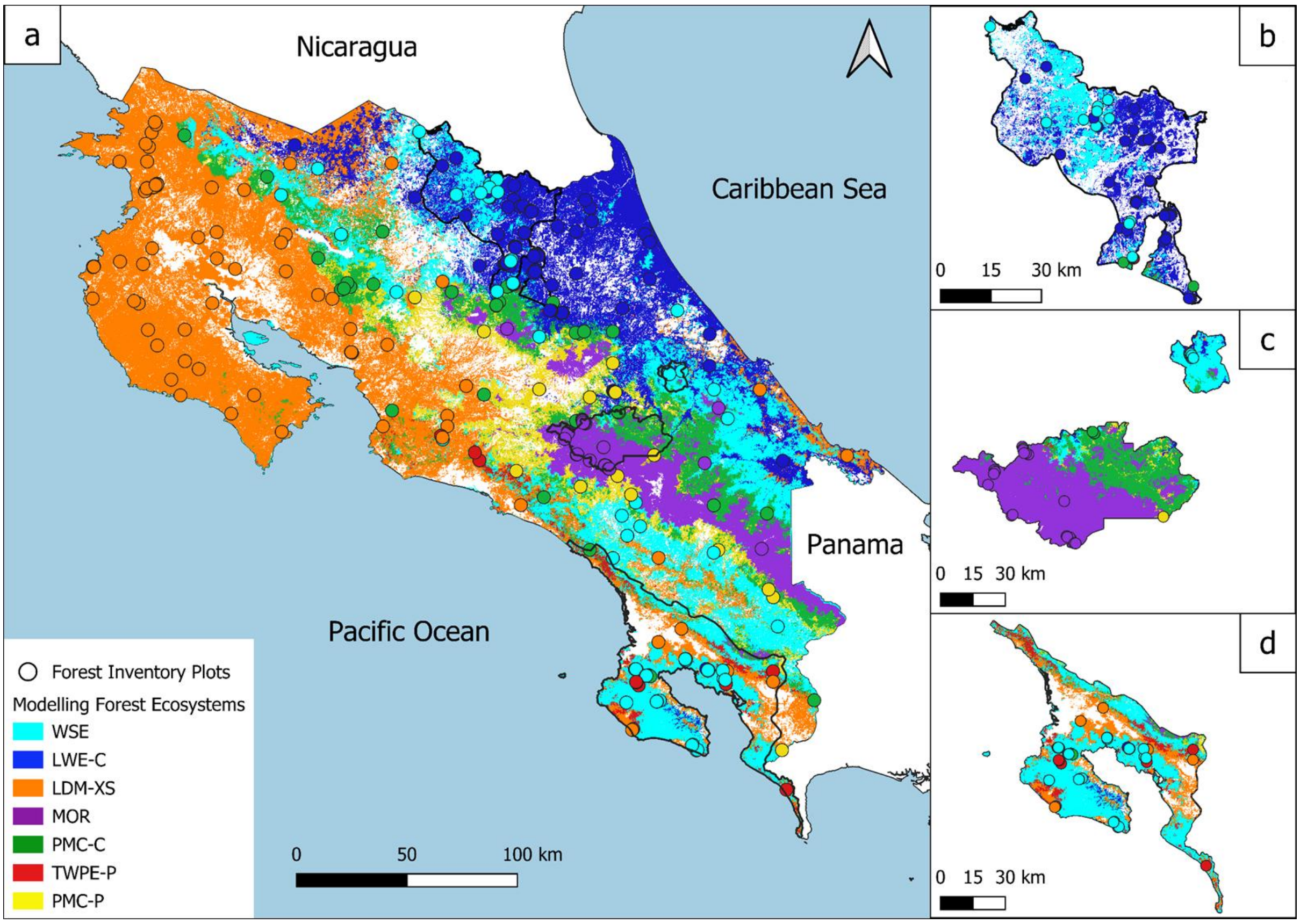

**Figure 6.** Map of Costarican forest ecosystems based on a classification model of seven main clusters, derived from dissimilarity between inventoried forest sites in terms of dominant tree species composition, using a random forest model. WSE: Wet seasonal evergreen forest. LWE-C: Lowland wet evergreen forest of Caribbean slope. LDM-DS: Lowland dry-to-moist deciduous-to-semi-deciduous forest. MOR: Mountain oak rainforest. PMC-C: Premontane-to-mountain mixed-to-evergreen cloud forest of Caribbean slope. TWPE-P: Transitional wet premontane evergreen forest of the Pacific slope. PMC-P: Premontane-to-mountain mixed-to-evergreen cloud forest of Pacific slope. Table provides the names of ecosystems and their acronyms. a, b, and c represent three areas where forests were characterized at the local scale, and the results will be interpreted in the discussion section.

### 3.4. Analysis of the distribution of forest-types within predicted forest ecosystems

The analysis of the distribution of OGF and SF across the clusters and derived forest ecosystems revealed notable inversions in the proportions of OGF and SF for ecosystems LWE-C and PMC-P between the results of the two methods (Table 5). Overall, all main clusters and associated forest ecosystems exhibit a dominance of one forest-type, with an average difference in OGF and SF proportions of 63±23% in the clustering results and 58±22% in those from Random Forest. A mean difference of 26 ± 22% in the proportions of forest types is observed between the clustering approach and the Random Forest approach, with this variation primarily due to the inverse predictions for two ecosystems, LWE-C and PMC-P.

**Table 5.** Comparison of the proportions of forest-types (Old-Growth Forests - OGF and Secondary Forests - SF) derived from the clustering of forest inventory plots and Random Forest modelling across the entire forest cover.

| Ecosystems | Forest Types | Proportion (%) | |
|---|---|---|---|
| | | Clustering | Random Forest |
| WSE | OGF | 0.79 | 0.69 |
| | SF | 0.21 | 0.31 |
| LWE-C | OGF | 0.74 | 0.25 |
| | SF | 0.26 | 0.75 |
| LDM-DS | OGF | 0.23 | 0.14 |
| | SF | 0.77 | 0.86 |
| MOR | OGF | 1 | 0.95 |
| | SF | 0 | 0.05 |
| PMC-C | OGF | 0.66 | 0.85 |
| | SF | 0.34 | 0.15 |
| TWPE-P | OGF | 0.93 | 0.63 |
| | SF | 0.07 | 0.37 |
| PMC-P | OGF | 0.18 | 0.79 |
| | SF | 0.82 | 0.21 |

## 4. Discussion

The objective of this study was to propose a replicable national approach to evaluate the potential of SF in maintaining dominant tree species structuring the forest ecosystems. Our approach was applied to Costa Rica. For that, we analysed dominant canopy species assemblages to better understand their contribution to the preservation and dynamics of forest ecosystems. Based on 364 plots, we identified and modelled seven major forest ecosystems in Costa Rica. These results relied on an approach that combined accessible data with local expertise. By integrating free environment geodata sources, such as satellite spectral information, global spatial environmental databases, and open-access digital elevation model (DEM), we produced a consistent forest mapping. Our approach achieved robust performance with an overall F1-score of 0.73 and a macro F1-score of 0.58, demonstrating the reliability of the model despite the imbalance between clusters. The challenge of differing spatial resolutions between floristic and environmental data was addressed using a segmentation method, dividing forest cover into spectrally homogeneous segments. This oriented objects approach was completed by statistical zoning based on percentiles, effectively capturing ecological trends specific to each segment and ensuring better representation of forest variations. The integration of expert knowledge was crucial for interpreting these data, refining models, and validating clusters through field campaigns. This method offers the advantage of being highly replicable; any country with local expertise and national forest inventory data can adopt it. Furthermore, the use of open-source software such as R and QGIS enhances its accessibility and global applicability.

### 4.1. Defining Forest Ecosystems through dominant tree Assemblages

The delineation of forest ecosystems was carried out using higher-level clusters to ensure a good balance between sampling size and satisfactory modeling performance across the entire forest cover. The classification performance of clusters was influenced by sampling biases in terms of accessibility and altitude (Table S6), with only 4% of plots between 1000–1500 m and 14% between 500–1000 m. This primarily affected the classification of ecosystems PMC-C, TWPE-P, and PMC-P within these ranges. Despite these limitations, the overall forest-type characterisation remains robust, supported by local-scale studies. Three well-documented sites in Costa Rica, located in ecologically diverse areas, confirm this consistency and demonstrate the effectiveness of our approach in complex contexts.

In the San Juan-La Selva Biological Corridor in northeastern Costa Rica, two forest ecosystems (WSE, and LWE-C) coexist (Figure 6b). Three lowland forest-types have been identified (Sesnie et al. 2008): *Pentaclethra macroloba* forests, dominant in LWE-C; *Qualea paraensis, Vochysia ferruginea*, and *Couma macrocarpa* forests, characteristic of WSE; and Dialium guianense, *Brosimum alicastrum*, and *Tachigali costaricensis* forests, spanning WSE and LWE-C. Additionally, piedmont and premontane species like *Vochysia allenii* and *Macrohasseltia macroterantha* are abundant in WSE.

In the rainforests of the Talamanca Cordillera, where ecosystems WSE, MOR, and PMC-C coexist along an altitudinal gradient up to 2,520 m (Figure 6c), three main forest-types have been characterised (Veintimilla et al. 2019): Lowland forests (440–1,120 m) are dominated by *Pourouma bicolor, Vochysia allenii*, and *Calophyllum brasiliense* (WSE); intermediate forests (1,400–1,660 m) host both lowland and montane species, primarily *Oreomunnea mexicana* (MOR), alongside *Billia rosea* (MOR), *Alchornea latifolia*, and *Cecropia insignis* (PMC-C); montane forests (2,150–2,950 m) define MOR, dominated by *Quercus bumeliodes, Drimys granadensis, Ocotea austinii*, and *Weinmannia pinnata* (MOR).

The Osa Peninsula, home to WSE, LWE-C, LDM-DS, TWPE-P, and PMC-P (Figure 6d), is among Costa Rica's most biodiverse regions due to its environmental and topographical variability (Hofhansl et al. 2020). With 28 forest-types, no single canopy species dominates (Gilbert & Kappelle 2016). Despite high endemism and wet forest dominance, lowland forests also include species from drier and moister habitats of Nicoya and the Central Valley, explaining LDM-DS continuity into the peninsula (Zamora et al. 2004). Some moist and wet lowland forest species of Osa peninsula also occur on the Caribbean coast, indicating a floristic and climatic affinity across the Talamanca Range, consistent with WSE, LWE-C, and TWPE-P distributions (Table S7).

The analysis of these examples shows that the local heterogeneity of forests is well represented by forest ecosystems based on dominant canopy species. It highlights the key role of oligarchic species in structuring assemblages across diverse, sometimes discontinuous, regions within the same ecosystem type. Specifically, *Vochysia spp.* and *Pourouma bicolor* define the WSE ecosystem, *Goethalsia meiantha* characterizes LWE-C, and *Cordia alliodora, Spondias spp., and Brosimum spp*. dominate LDM-DS (Sesnie et al. 2008; Gilbert & Kappelle 2016). The distribution of oligarchic species is strongly correlated with geographical and topographical variables, particularly in the lowland forests of the Osa Peninsula (Hofhansl et al. 2019; Morera-Beita et al. 2019). These findings enhance the understanding of the regional spatial complex distribution of ecosystems at large spatial scales (Figure 6). The contribution values of species are available in Table S2.

## 4.2. Importance of spectral and environmental factors in ecosystem characterisation and their interactions

Our approach identifies contributive species assemblages that significantly define forest ecosystems, enabling analysis of their interactions and specificities. Using abundance data from forest inventories, we calculate the average fraction of species contributing to cluster j that also contribute to cluster j' (Lenormand et al. 2019). In Costa Rica, assemblages show greater specificity in lowland (WSE, LWE-C, LDM-DS) and in high-mountain (MOR) forests but are less distinct in intermediate ecosystems (PMC-C, TWPE-P, PMC-P). Low interaction rates between ecosystems at the national level reflect abiotic filters shaping habitat diversity, which drives high biodiversity (Araújo & Rozenfeld 2014).

Forest ecosystems, mostly, are structured by environmental gradients that directly influence the contribution of dominant canopy species, which have played a key role in their definition. Through Random Forest modelling, we observe that topography, followed by climate and the NDWI index—used here to estimate water content in vegetation and, consequently, water stress—are the most determining factors in the turnover of tree species composition that structure ecosystems. However, edaphic properties also play an important role, as evidenced by the selection of pH and CEC among the seven variables chosen for the modelling. Our results are consistent with previous studies conducted in tropical forests across South and Central America, where tree species turnover has been associated with these environmental variables, as well as canopy reflectance, at spatial scales ranging from regional to continental (e.g., Pérez Chaves et al. 2020; Bañares-de-Dios et al. 2022; Jakovac et al. 2022; Nuñez et al. 2024). The integration of these environmental determinants with the analysis of contributive species assemblages, allowed for the characterisation and naming of forest ecosystems derived from in situ clusters (Table 4). This designation is based on two national, well established, ecological-vegetation mapping systems: life zones (Holdridge 1967), defined by bioclimatic factors, and phytogeographic units (Zamora 2008), determined by floristic composition associated with topographic, climatic, and soil type factors. A detailed description of these ecosystems, integrating the contribution of key environmental variables in the modelling as well as the assemblages of species that significantly characterise them, is provided in Table S8**.**

## 4.3. Contribution of secondary forests to the dynamics and conservation of national forest ecosystems

Anthropogenic pressures shape forest ecosystems by influencing species composition, distribution and interactions. The spread of SF, driven by past and present disturbances (Arroyo-Rodríguez et al. 2017), helps assess human impact. While SF distribution differs minimally between in situ clusters and model-derived ecosystems, LWE-C and PMC-P show inverse trends: modelling predicts more SF in LWE-C and more OGF in PMC-P, contrasting with clustering results obtained from in situ data. This may stem from sampling bias favouring OGF plots in accessible, lower-altitude areas, particularly in LWE-C. However, for PMC-P, botanical expertise and national disturbance records confirm its accuracy. In Costa Rica, colonization history has influenced land use, shaping the distribution of OGF and SF (Redo et al. 2012; Aide et al. 2013; Shaver et al. 2015). Since the pre-Columbian times, human activity on the northern Pacific coast and central highlands (LDM-DS) converted dry forests mainly into cattle pastures, later abandoned and replaced by SF mosaics. During colonial times, deforestation expanded to pre-mountain Pacific areas (PMC-P) due to their proximity to the central valley, mild climate, and fertile soils, ideal for coffee cultivation, which fuelled SF expansion. On the Caribbean coast (LWE-C), intense deforestation (1960s–1980s) in particular for banana and pineapple monocultures fragmented OGF, restricting it to inaccessible high-altitude zones (PMC-C, MOR). In Osa Peninsula lowlands, about half the forest remains OGF (WSE and TWPE-P), though degradation from logging and oil palm plantations that has driven SF expansion in LWE-C, LDM-DS, and TWPE-P. These findings align with Random Forest modelling, which corrected OGF bias in LWE-C using ecological variables. However, for PMC-P, model limitations persisted, reflected in a low F1-score (0.38). The SF proportion in each ecosystem reflects the level of anthropogenic pressure reshaping the forested landscape, playing a crucial role in floristic dynamics and significantly influencing the evolution of forest floras.

Analysis of SF contributions to forest ecosystems revealed their limited ability to maintain OGF-specific tree assemblages, identified through clustering and Random Forest modelling. In Costa Rica, OGF dominates WSE, MOR, PMC-C, and TWPE-P, while SF prevails in LWE-C, LDM-DS, and PMC-P. This suggests SF, in their current stages, remain compositionally distinct from OGF (Rozendaal et al. 2019; Mertz et al. 2021) especially in MOR, where OGF fully dominates due to inaccessibility. However, SF ecosystems like LWE-C and LDM-DS should not be seen as degraded but as distinct ecosystems with their own species assemblages (Pain et al. 2021). OGF and SF distinctions blur at the canopy level, as both share many species, though one tends to dominate per ecosystem. Some plots classified as OGF are likely forests that have undergone a certain degree of human intervention, which explains the significant presence of species typical of SF or fast-growing species. In OGF-dominated landscapes with low-intensity land use, SF could follow optimal successional paths, helping preserve native flora (Rosenfield et al. 2023).

This study provides a first assessment of forest ecosystem vulnerability at the national level. In the case of Costa Rica, MOR appears the most threatened, with its contributive species strictly confined to high altitudes (>2150 m) and a 90% specificity rate. Its high vulnerability arises from SF's inability to sustain it and the potential loss of its ecological niche due to climate change, despite low anthropogenic pressure. In contrast, PMC-C seems the least vulnerable, dominated by OGF with an 85% specificity rate and benefiting from natural protection due to inaccessibility. Its species adapt to a wider altitudinal range (<1300 m), enhancing resilience to climate change. Lowland and mid-mountain ecosystems (WSE, LWE-C, LDM-DS, TWPE-P, PMC-P) face high anthropogenic pressures, with specificity rates of 78%-99%. In these areas, young SF are particularly vulnerable, often seen as fallow lands (Reid et al. 2019), threatening their regeneration potential and long-term ecosystem stability.

### 4.4. Limits and perspectives

The approach used has limitations related to the quantity and distribution of available field data. With only 364 sampling plots at the national scale, the classification of forest ecosystems in Costa Rica faced a sampling power imbalance between clusters, impacting the modelling performance in particular for clusters C5 to C7. This imbalance is partly due to sampling biases related to the accessibility and altitudinal distribution of the plots, favouring forests located at low and medium altitudes. Moreover, the variability of the plots by forest-types (OGF and SF) in the sampling also affects the species composition in each assemblage. Classification effectiveness depends on the quantity and representativeness of available data. Well-sampled clusters (C1–C3) show clear ecological trends and high performance, while underrepresented clusters (C5–C7) have broader, harder-to-distinguish trends. Cluster C4 is an exception—despite being under-sampled, its strict ecological characteristics enable better classification. In Costa Rica, these findings underscore the need for greater financial investment in forest inventories to enhance model accuracy by increasing plot numbers and improving representation of mid- and high-altitude forests, which are harder to access.

Data acquisition remains a major challenge, highlighting countries' difficulties in collecting essential biodiversity data for monitoring, tracking, and modelling. Significant disparities exist in generating in situ data due to high costs, lack of standardization, and coordination challenges. These limitations hinder comprehensive data collection, affecting national biodiversity strategies and global assessments (Chapman et al. 2024). As a result, global biodiversity data remains highly uneven, hindering accurate assessments of ecosystem status, extent, and distribution (Gonzalez & Londoño 2022). This data gap weakens efforts to guide conservation actions and monitor progress toward protecting 30% of land by 2030, a core target of the Kunming-Montreal Global Biodiversity Framework (GBF) adopted in 2022 (CBD 2022). The GBF emphasizes ecosystem integrity and connectivity, yet lacks explicit targets for national biodiversity monitoring systems (Perino et al. 2022). The Ecosystem Extent Indicator, a key metric for tracking ecosystem loss and degradation at the global level, highlights the urgency of strengthening standardized monitoring frameworks to assess ecosystem trends effectively. However, disparities in data availability, financial resources, and technical capacity create major obstacles for many countries. International financial and technical support is therefore essential to enable comprehensive biodiversity monitoring, ecosystem extent assessments, and effective conservation planning (Gonzalez & Londoño 2022; Cardona Santos et al. 2023). Without these investments, achieving the GBF 2030 targets and reversing biodiversity loss will remain a significant challenge. In the absence of such funding, the deployment of cost-effective monitoring systems relies on integrating remote sensing with biodiversity surveillance models at national scales, complementing in situ data to enhance spatial representativeness (Fernández et al. 2020). These large-scale extrapolated quantitative models must be incorporated into participatory approaches that engage both expert and local knowledge to strengthen the effective implementation of the GBF at the national level (Rosa et al. 2017; Xu et al. 2021).

In this context, the approach developed and tested in this study—designed to be operational and replicable in other tropical countries—aims to enhance the identification and assessment of tropical forest ecosystems by analyzing the contribution of SF. Importantly, this operational framework represents a critical step toward prioritizing tropical forest conservation at the national level, directly supporting the goal of protecting 30% of terrestrial land. By integrating national forest inventories, spectral data, and global environmental databases, the approach enables the identification and characterization of forest ecosystems based on contributive species assemblages, ecosystem interactions, and ecological specificity. Moreover, it provides key insights into the role of SF in maintaining and shaping forest dynamics. Besides, the network of interactions (Figure 5) highlights the specificity of different clusters or bioregions in terms of species assemblages and their interrelationships. This approach allows rapid identification of highly specific clusters as well as transition areas, providing a concise summary of ecological relationships across the country. While the primary focus of this study is on mapping and characterizing species assemblages, these network patterns offer valuable insights for future conservation planning. In particular, key network metrics (e.g., $\lambda_{jj'}$) could help identify priority areas and guide biodiversity management at regional and national scales, representing an important avenue for subsequent research.

From a conservation prioritization perspective, developing a vulnerability categorization for ecosystems facing anthropogenic pressures and climate change (Guariguata & Ostertag 2001; Edwards et al. 2019) is essential. This can be achieved using vulnerability indices (Kumar et al. 2021; Roshani et al. 2024) that incorporate ecosystem specificity, interaction strength, and detailed analyses of how contributive species respond to disturbances (Fremout et al. 2020; Pang et al. 2023). Future research should aim to explicitly link the identified tree assemblages to key ecosystem functions and biodiversity metrics (Lohbeck et al. 2016; Fichtner & Härdtle 2021; Zhang & Zang 2021; Cooper et al. 2024). Such analyses would help clarify whether different assemblages contribute in complementary or redundant ways to ecosystem multifunctionality, thereby strengthening the conservation implications of the delineated forest ecosystems. To further reduce sampling biases, the integration of citizen science data, such as those available through GBIF, should be considered in future studies (Fraisl et al. 2022). Additionally, a fully remote sensing-based approach could be employed to generate spectral clusters that capture spectral diversity, which may strongly correlate with species richness, beta diversity, and functional diversity derived from in situ observations (Féret & de Boissieu 2020; Chraibi et al. 2021; Perrone et al. 2023; Lenormand et al. 2025). These perspectives represents a first step towards improving understanding of the role of secondary forests in conserving tropical forest ecosystems at the national scale.

## CRediT

**Maïri Souza Oliveira:** Conceptualization, Formal analysis, Data curation, Formal analysis, Validation, Investigation, Visualization, Writing – original draft, Methodology, Writing – review and editing. **Maxime Lenormand:** Software, Investigation, Methodology, Writing – review and editing. **Sandra Luque:** Conceptualization, Funding acquisition, Supervision, Project administration, Investigation, Visualization, Methodology, Writing – review and editing. **Nelson A. Zamora:** Conceptualization, Expert Validation, Investigation, Methodology, Writing – review and editing, Resources. **Samuel Alleaume :** Resources, Methodology, Writing – review and editing. **Adriana C. Aguilar Porras:** Resources, Investigation. **Marvin U. Castillo:** Resources. **Eduardo Chacón-Madrigal:** Resources, Writing – review and editing. **Diego Delgado:** Resources, Investigation. **Luis Gustavo Hernández Sánchez:** Resources. **Marie-Ange Ngo Bieng:** Conceptualization, Funding acquisition, Supervision, Investigation, Writing – review and editing. **Ruperto M. Quesada:** Resources. **Gilberth S. Solano:** Resources. **Pedro M. Zúñiga:** Resources.

## Acknowledgements

This research was partially funded by a PhD fellowship from INRAE and CIRAD, with additional support from CNES (National Space Agency, France) TOSCA Project CODEX for fieldwork and data validation. We sincerely thank SINAC, CATIE, ITCR, Fundecor, UNA, and CODEFORSA for granting access to forest inventory databases. We are grateful to Mona Bonier, Florian de Boissieu, Raffaele Gaetano, Cassio Fraga Dantas, Jean-Baptiste Féret, and Dino Ienco (TETIS) for their valuable support in various aspects of data analysis and expertise. Thanks also to Sergio Vilchez-Mendoza (CATIE) for his advice and to Daniel Villavicencio (SINAC) and Eliott Maurent (CIRAD) for providing the data and explanations, which were essential to the project. Special thanks to CATIE and SINAC for their hospitality during fieldwork, Luc Villain (CIRAD) for administrative support, and the SINAC park guards and technicians and Vicente Herra for their essential field assistance during our in situ validation campaign.

## Declaration of competing interest

The authors declare that they have no known competing financial interests or personal relationships that could have appeared to influence the work reported in this paper.

## Data Availability Statement

The in situ data used in this study are the property of Costa Rican institutions and were made available exclusively for the purposes of this research. For any requests regarding access to these data, please contact the co-authors affiliated with the relevant Costa Rican institutions. All other data used to develop the various variables, as well as the code and data necessary to assess the study's conclusions, are referenced in the manuscript or included in the supplementary material.

## Supplementary material

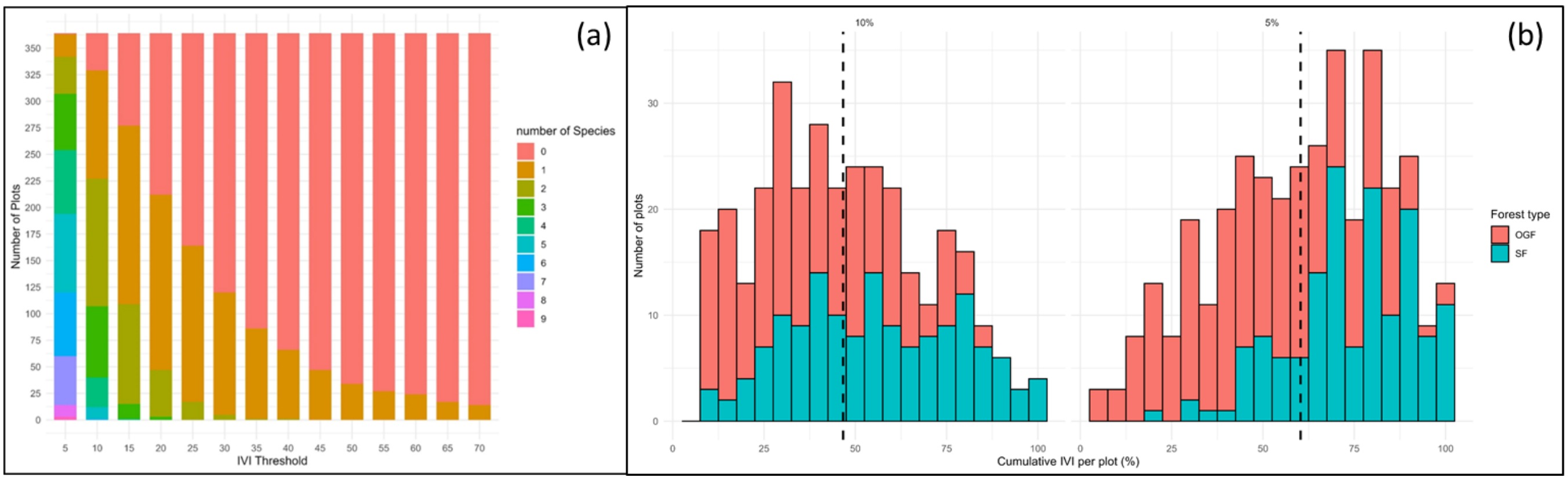

**Figure S1.** (a) Fraction of plots with a given number of dominant tree species according to the Importance Value Index (IVI) threshold ; and (b) Distribution of cumulative IVI captured by dominant species according to two thresholds, 5 % for the barplot on the right and 10 % for the one on the left. The y-axis represents cumulative IVI (%), and the x-axis the number of forest plots. Each bar is segmented to show the number of plots by forest type (SF – secondary forest, OGF – old-growth forest). The dashed line indicates the mean cumulative IVI for each threshold.

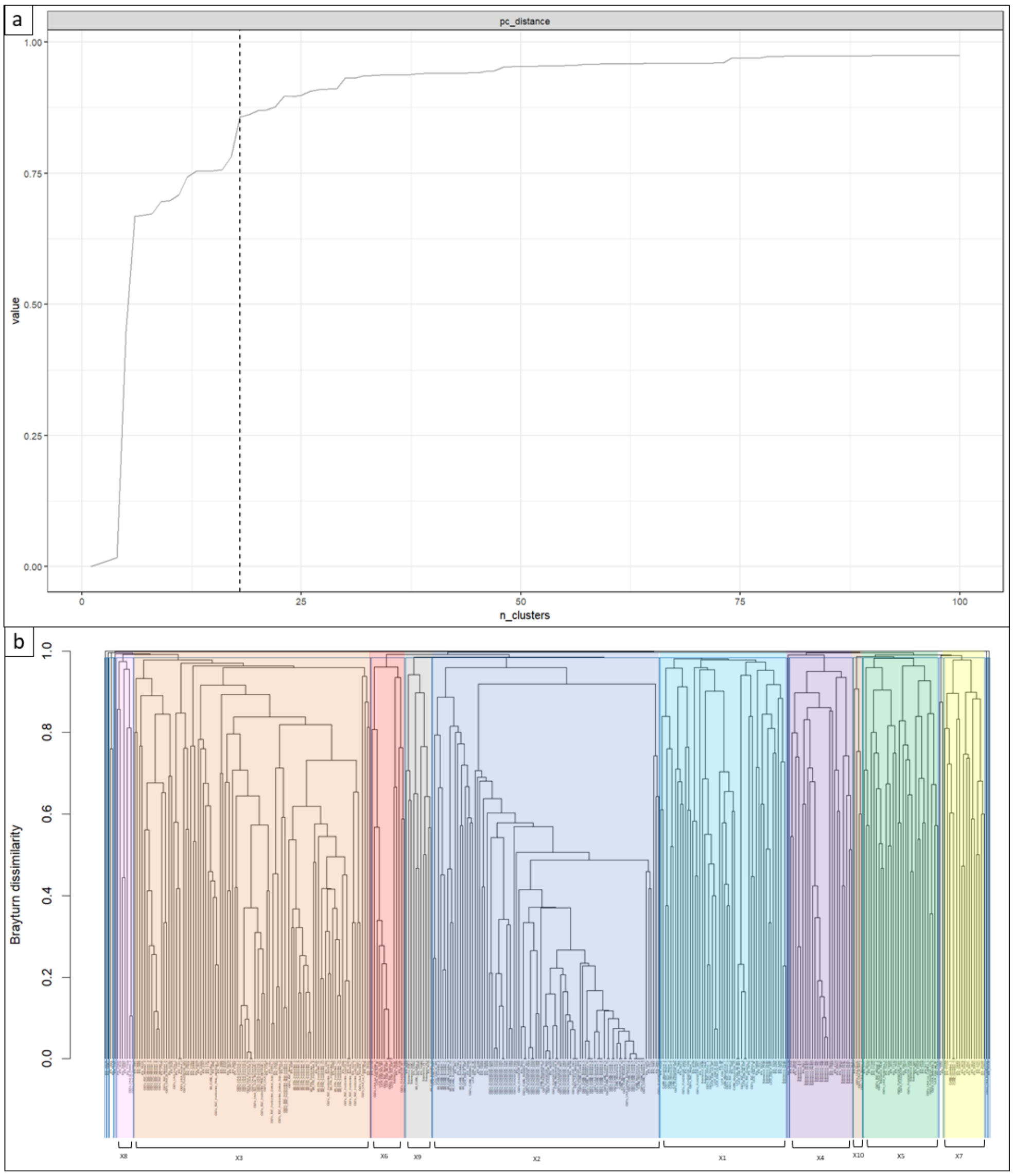

**Figure S2.** Results of hierarchical clustering based on the dissimilarity index Bray-Curtis Turnover and UPGMA method. (a) Explained dissimilarity by different number of clusters with the potential optimal number of clusters based on the use of the pc_distance metric in the elbow method, and (b) Hierarchical clustering of 364 sites with the top 10 clusters selected for biogeographical network analysis. The optimal PC_distance is 18, and the cophenetic correlation coefficient of the hierarchical clustering model is 0.48.

**Table S1.** Results of the assessment of isolated sites corresponding to cluster j using Simpson's distance coupled with Bonferroni correction: To distinguish isolated sites with a real floristic specificity from those that simply exhibit a high rate of unique species due to random sampling, a pairwise similarity test was conducted between the isolated sites and the main clusters, using Simpson's distance coupled with a Bonferroni correction. Based on botanical expertise, a similarity threshold of 50% was chosen to allow the association of isolated sites with a similar floristic composition to the main clusters.

| Cluster j | Cluster j' | Simpson |
|---|---|---|
| C11 | C6 | 0.58 |
| | C1 | 0.46 |
| | C2 | 0.29 |
| | C5 | 0.21 |
| | C9 | 0.17 |
| | C3 | 0.17 |
| | C8 | 0.17 |
| | C7 | 0.13 |
| C12 | C3 | 0.25 |
| | C7 | 0.25 |
| | C8 | 0.25 |
| C13 | C1 | 0.38 |
| | C5 | 0.33 |
| | C8 | 0.33 |
| | C6 | 0.30 |
| | C4 | 0.20 |
| | C9 | 0.18 |
| | C7 | 0.18 |
| | C2 | 0.15 |
| | C3 | 0.05 |
| C14 | C3 | 0.56 |
| | C2 | 0.11 |
| | C6 | 0.11 |
| C15 | C3 | 0.78 |
| | C8 | 0.48 |
| | C6 | 0.26 |
| | C1 | 0.17 |
| | C2 | 0.09 |
| | C5 | 0.04 |
| C16 | C6 | 0.41 |
| | C9 | 0.29 |
| | C3 | 0.29 |
| | C1 | 0.29 |
| | C2 | 0.24 |
| | C5 | 0.18 |
| | C7 | 0.06 |
| C17 | C5 | 0.65 |
| | C10 | 0.51 |
| | C8 | 0.21 |
| | C4 | 0.21 |
| | C9 | 0.16 |
| | C6 | 0.15 |
| | C1 | 0.14 |
| | C2 | 0.13 |
| | C7 | 0.13 |
| | C3 | 0.02 |
| C18 | C3 | 1.00 |
| | C1 | 0.33 |
| | C6 | 0.33 |

**Table S2.** List of contributive species ($\rho_{ij}$ ≥ 1.96) by in situ cluster.

| cluster | species | pij |
|---|---|---|
| C1 | Elaeoluma glabrescens | 9.65 |
| | Symphonia globulifera | 8.68 |
| | Garcinia madruno | 8.39 |
| | Ruptiliocarpon caracolito | 7.55 |
| | Compsoneura excelsa | 7.34 |
| | Humiriastrum diguense | 6.89 |
| | Brosimum guianense | 6.87 |
| | Cassipourea elliptica | 6.87 |
| | Tapirira guianensis | 6.83 |
| | Mabea occidentalis | 6.81 |
| | Couratari guianensis | 6.47 |
| | Protium stevensonii | 6.40 |
| | Vochysia ferruginea | 6.39 |
| | Pouteria laevigata | 6.34 |
| | Brosimum lactescens | 6.11 |
| | Vochysia allenii | 5.85 |
| | Virola koschnyi | 5.75 |
| | Calophyllum mesoamericanum | 5.58 |
| | Peltogyne purpurea | 5.55 |
| | Perebea hispidula | 5.53 |
| | Dialium guianense | 5.44 |
| | Eschweilera calyculata | 5.42 |
| | Manilkara staminodella | 5.40 |
| | Inga thibaudiana | 5.38 |
| | Vantanea barbourii | 5.37 |
| | Pera arborea | 5.36 |
| | Pouteria torta | 4.94 |
| | Calophyllum brasiliense | 4.93 |
| | Trichilia septentrionalis | 4.92 |
| | Qualea paraensis | 4.79 |
| | Matudaea trinervia | 4.60 |
| | Dendropanax arboreus | 4.43 |
| | Xylopia sericophylla | 4.38 |
| | Marila laxiflora | 4.31 |
| | Miconia poeppigii | 4.31 |
| | Virola nobilis | 4.25 |
| | Caryocar costaricense | 4.19 |
| | Vochysia gentryi | 4.01 |
| | Apeiba tibourbou | 3.97 |
| | Metteniusa tessmanniana | 3.84 |
| | Brosimum utile | 3.84 |
| | Carapa nicaraguensis | 3.74 |
| | Copaifera camibar | 3.70 |
| | Pithecellobium dulce | 3.57 |
| | Micropholis crotonoides | 3.39 |
| | Couma macrocarpa | 3.33 |
| | Tovomita weddelliana | 3.27 |
| | Castilla tunu | 3.22 |
| | Swartzia simplex | 3.21 |
| | Miconia affinis | 3.19 |
| | Cupania dentata | 3.12 |
| | Elvasia elvasioides | 3.11 |
| | Dussia macroprophyllata | 3.04 |
| | Hirtella triandra subsp. media | 2.94 |
| | Guarea rhopalocarpa | 2.90 |
| | Sloanea medusula | 2.90 |
| | Sorocea pubivena | 2.85 |
| | Virola sebifera | 2.83 |
| | Tetrathylacium macrophyllum | 2.80 |
| | Virola montana | 2.71 |
| | Trichospermum galeottii | 2.62 |
| | Simarouba amara | 2.57 |
| | Eschweilera biflava | 2.54 |
| | Jacaranda copaia | 2.51 |
| | Macrohasseltia macroterantha | 2.51 |
| | Inga venusta | 2.48 |
| | Hieronyma alchorneoides | 2.46 |
| | Cupania glabra | 2.39 |
| | Vochysia guatemalensis | 2.39 |
| | Pouteria chiricana | 2.32 |
| | Henriettea fascicularis | 2.32 |
| | Ficus tuerckheimii | 2.32 |
| | Axinaea costaricensis | 2.32 |
| | Lonchocarpus sericeus | 2.32 |
| | Calliandra trinervia | 2.32 |
| | Piptocoma discolor | 2.32 |
| | Bellucia grossularioides | 2.32 |
| | Ficus velutina | 2.26 |
| | Anaxagorea crassipetala | 2.24 |
| | Inga tenuiloba | 2.23 |
| | Maranthes panamensis | 2.20 |
| | Guatteria ucayalina | 2.11 |
| | Chimarrhis latifolia | 2.08 |
| | Rinorea dasyadena | 2.08 |
| | Miconia bigibbosa | 2.07 |
| | Xylopia sericea | 2.03 |
| | Pourouma bicolor | 1.99 |
| | Hieronyma oblonga | 1.98 |

| cluster | species | pij |
|---|---|---|
| C2 | Pentaclethra macroloba | 11.62 |
| | Matisia pacifica | 8.24 |
| | Hernandia didymantha | 7.32 |
| | Protium confusum | 7.04 |
| | Stryphnodendron microstachyum | 7.04 |
| | Protium panamense | 7.03 |
| | Virola sebifera | 7.03 |
| | Apeiba membranacea | 6.98 |
| | Minquartia guianensis | 6.89 |
| | Cordia dwyeri | 6.87 |
| | Protium pittieri | 6.72 |
| | Ocotea laetevirens | 6.51 |
| | Pourouma bicolor | 6.35 |
| | Sacoglottis trichogyna | 6.22 |
| | Dendropanax arboreus | 6.21 |
| | Pourouma minor | 6.20 |
| | Protium ravenii | 6.13 |
| | Brosimum lactescens | 5.88 |
| | Warszewiczia coccinea | 5.67 |
| | Chrysophyllum colombianum | 5.48 |
| | Morisonia pittieri | 5.29 |
| | Casearia arborea | 5.26 |
| | Inga laevigata | 5.08 |
| | Maranthes panamensis | 5.04 |
| | Byrsonima arthropoda | 4.90 |
| | Casearia corymbosa | 4.73 |
| | Clethra costaricensis | 4.66 |
| | Alchorneopsis floribunda | 4.61 |
| | Conceveiba pleiostemona | 4.56 |
| | Goethalsia meiantha | 4.51 |
| | Dipteryx panamensis | 4.47 |
| | Carapa guianensis | 4.44 |
| | Casearia bicolor | 4.37 |
| | Guatteria aeruginosa | 4.32 |
| | Rhodostemonodaphne kunthiana | 4.32 |
| | Annona papilionella | 4.31 |
| | Tachigali costaricensis | 4.22 |
| | Balizia elegans | 4.19 |
| | Jacaratia spinosa | 4.01 |
| | Miconia elata | 3.73 |
| | Ferdinandusa panamensis | 3.63 |
| | Ilex fortunensis | 3.58 |
| | Hirtella triandra | 3.58 |
| | Simarouba amara | 3.50 |
| | Croton smithianus | 3.29 |
| | Cordia bicolor | 3.28 |
| | Bunchosia argentea | 3.22 |
| | Terminalia amazonia | 3.16 |
| | Dystovomita paniculata | 3.09 |
| | Lacistema aggregatum | 3.08 |
| | Inga pezizifera | 3.06 |
| | Hasseltia floribunda | 3.03 |
| | Annona montana | 3.02 |
| | Celtis schippii | 3.02 |
| | Inga thibaudiana | 3.01 |
| | Tabernaemontana donnell-smithii | 3.00 |
| | Brosimum guianense | 3.00 |
| | Vitex cooperi | 2.93 |
| | Tapirira guianensis | 2.80 |
| | Miconia multispicata | 2.80 |
| | Simira maxonii | 2.78 |
| | Unonopsis pittieri | 2.68 |
| | Pterocarpus rohrii | 2.65 |
| | Posoqueria latifolia | 2.60 |
| | Lonchocarpus ferrugineus | 2.57 |
| | Pterocarpus officinalis | 2.53 |
| | Ocotea rivularis | 2.52 |
| | Pouteria torta | 2.50 |
| | Brosimum lactescens | 2.45 |
| | Virola koschnyi | 2.38 |
| | Dussia macroprophyllata | 2.35 |
| | Croton schiedeanus | 2.33 |
| | Anaxagorea crassipetala | 2.31 |
| | Xylopia sericophylla | 2.31 |
| | Vismia macrophylla | 2.30 |
| | Graffenrieda galeottii | 2.22 |
| | Castilla elastica | 2.22 |
| | Ocotea atirrensis | 2.21 |
| | Licaria misantlae | 2.14 |
| | Adelia triloba | 2.13 |
| | Cupania glabra | 2.11 |
| | Grias cauliflora | 2.11 |
| | Ocotea pullifolia | 2.11 |
| | Cordia cymosa | 2.05 |

| cluster | species | pij |
|---|---|---|
| C3 | Handroanthus ochraceus | 8.25 |
| | Bursera simaruba | 7.57 |
| | Spondias mombin | 7.55 |
| | Cordia alliodora | 6.93 |
| | Albizia niopoides | 6.91 |
| | Guazuma ulmifolia | 6.66 |
| | Sterculia apetala | 6.64 |
| | Cochlospermum vitifolium | 6.59 |
| | Luehea speciosa | 6.12 |
| | Maclura tinctoria | 6.07 |
| | Chomelia spinosa | 6.02 |
| | Lonchocarpus felipei | 5.92 |
| | Enterolobium cyclocarpum | 5.36 |
| | Calycophyllum candidissimum | 5.25 |
| | Myrospermum frutescens | 4.94 |
| | Eugenia hiraeifolia | 4.92 |
| | Cordia panamensis | 4.63 |
| | Brosimum alicastrum | 4.58 |
| | Samanea saman | 4.39 |
| | Sideroxylon capiri | 4.23 |
| | Cassia grandis | 4.22 |
| | Lonchocarpus rugosus | 4.17 |
| | Thouinidium decandrum | 4.17 |
| | Bravaisia integerrima | 4.15 |
| | Pochota fendleri | 3.84 |
| | Anacardium excelsum | 3.73 |
| | Luehea candida | 3.66 |
| | Tabernaemontana glabra | 3.65 |
| | Spondias purpurea | 3.64 |
| | Astronium graveolens | 3.64 |
| | Leptolobium panamense | 3.58 |
| | Trichilia americana | 3.56 |
| | Swietenia macrophylla | 3.53 |
| | Schizolobium parahyba | 3.53 |
| | Piscidia carthagenensis | 3.53 |
| | Quercus oleoides | 3.49 |
| | Triplaris melaenodendron | 3.49 |
| | Trichilia pleeana | 3.42 |
| | Diphysa americana | 3.41 |
| | Handroanthus impetiginosus | 3.40 |
| | Albizia adinocephala | 3.40 |
| | Lonchocarpus costaricensis | 3.40 |
| | Lonchocarpus minimiflorus | 3.23 |
| | Sloanea terniflora | 3.19 |
| | Hymenaea courbaril | 3.14 |
| | Gliricidia sepium | 3.13 |
| | Annona purpurea | 3.08 |
| | Simarouba glauca | 3.06 |
| | Trichilia martiana | 2.92 |
| | Solenandra mexicana | 2.86 |
| | Cenostigma eriostachys | 2.84 |
| | Inga vera | 2.81 |
| | Mespilodaphne veraguensis | 2.77 |
| | Gyrocarpus jatrophifolius | 2.77 |
| | Lysiloma auritum | 2.75 |
| | Lysiloma divaricatum | 2.74 |
| | Karwinskia calderonii | 2.70 |
| | Lonchocarpus parviflorus | 2.69 |
| | Guettarda macrosperma | 2.68 |
| | Manilkara chicle | 2.67 |
| | Aralia excelsa | 2.56 |
| | Cecropia peltata | 2.51 |
| | Casearia laetioides | 2.49 |
| | Luehea seemannii | 2.48 |
| | Rehdera trinervis | 2.42 |
| | Quararibea asterolepis | 2.41 |
| | Pterocarpus michelianus | 2.35 |
| | Sebastiania pavoniana | 2.28 |
| | Zanthoxylum setulosum | 2.24 |
| | Pseudosamanea guachapele | 2.23 |
| | Byrsonima crassifolia | 2.18 |
| | Lonchocarpus guatemalensis | 2.16 |
| | Bursera tomentosa | 2.10 |
| | Lonchocarpus macrophyllus | 2.06 |
| | Inga litoralis | 2.05 |
| | Ardisia revoluta | 2.03 |
| | Cupania guatemalensis | 2.00 |

| cluster | species | ρij |
|---|---|---|
| C4 | Quercus sapotifolia | 11.99 |
| | Weinmannia pinnata | 11.19 |
| | Drimys granadensis | 10.84 |
| | Ilex pallida | 10.00 |
| | Styrax argenteus | 9.89 |
| | Ocotea austinii | 9.79 |
| | Quercus costaricensis | 8.32 |
| | Cleyera theaeoides | 8.18 |
| | Billia rosea | 8.04 |
| | Quetzalia occidentalis | 7.86 |
| | Sciodaphyllum pittieri | 7.46 |
| | Symplocos serrulata | 7.44 |
| | Podocarpus oleifolius | 7.35 |
| | Oreomunnea mexicana | 7.35 |
| | Magnolia poasana | 7.24 |
| | Prumnopitys standleyi | 6.59 |
| | Miconia brevitheca | 6.50 |
| | Clethra consimilis | 6.48 |
| | Elaeagia glossostipula | 6.23 |
| | Hedyosmum bonplandianum | 5.57 |
| | Ladenbergia brenesii | 5.28 |
| | Dendropanax caucanus | 5.28 |
| | Miconia schnellii | 5.22 |
| | Brunellia costaricensis | 5.06 |
| | Escallonia myrtilloides | 4.87 |
| | Beilschmiedia alloiophylla | 4.58 |
| | Hieronyma oblonga | 4.26 |
| | Miconia tonduzii | 4.20 |
| | Ardisia pleurobotrya | 3.84 |
| | Verbesina oerstediana | 3.84 |
| | Hedyosmum goudotianum | 3.84 |
| | Myrsine coriacea | 3.67 |
| | Guatteria oliviformis | 3.52 |
| | Viburnum costaricanum | 3.30 |
| | Alfaroa costaricensis | 3.22 |
| | Miconia brenesii | 3.06 |
| | Quercus insignis | 2.91 |
| | Pleurothyrium palmanum | 2.63 |
| | Ilex lamprophylla | 2.52 |
| | Miconia durandii | 2.26 |

| cluster | species | ρij |
|---|---|---|
| C5 | Ruagea glabra | 6.46 |
| | Elaeagia auriculata | 6.15 |
| | Salacia petenensis | 6.14 |
| | Cecropia angustifolia | 5.67 |
| | Inga oerstediana | 5.58 |
| | Staphylea occidentalis | 5.26 |
| | Guarea kunthiana | 5.22 |
| | Dendropanax globosus | 5.13 |
| | Miconia conomicrantha | 5.11 |
| | Tetrorchidium euryphyllum | 5.10 |
| | Ocotea endresiana | 5.01 |
| | Vismia baccifera | 4.49 |
| | Chrysochlamys allenii | 4.42 |
| | Pterocarpus rohrii | 4.41 |
| | Lippia myriocephala | 4.41 |
| | Ficus crassiuscula | 4.41 |
| | Inga leonis | 4.35 |
| | Ardisia palmana | 4.30 |
| | Sapium rigidifolium | 4.05 |
| | Panopsis costaricensis | 3.89 |
| | Miconia conorufescens | 3.87 |
| | Colubrina spinosa | 3.81 |
| | Croton draco | 3.78 |
| | Miconia lasiopoda | 3.78 |
| | Aegiphila anomala | 3.63 |
| | Meliosma occidentalis | 3.63 |
| | Meliosma vernicosa | 3.48 |
| | Talisia macrophylla | 3.40 |
| | Warszewiczia uxpanapensis | 3.34 |
| | Senna papillosa | 3.33 |
| | Hampea appendiculata | 3.30 |
| | Erythrina steyermarkii | 3.27 |
| | Miconia biperulifera | 3.27 |
| | Alchornea glandulosa | 3.24 |
| | Ocotea macrophylla | 3.20 |
| | Casearia tacanensis | 3.19 |
| | Heliocarpus americanus | 3.16 |
| | Miconia brenesii | 3.06 |
| | Miconia prasina | 2.95 |
| | Guarea guidonia | 2.95 |
| | Miconia durandii | 2.84 |
| | Platymiscium curuense | 2.81 |
| | Guarea tonduzii | 2.79 |
| | Alchornea latifolia | 2.70 |
| | Coccoloba tuerckheimii | 2.63 |
| | Ilex lamprophylla | 2.56 |
| | Lonchocarpus monteviridis | 2.43 |
| | Ocotea stenoneura | 2.39 |
| | Persea caerulea | 2.39 |
| | Guarea glabra | 2.37 |
| | Inga barbourii | 2.30 |
| | Cecropia insignis | 2.30 |
| | Terminalia oblonga | 2.30 |
| | Pleurothyrium palmanum | 2.14 |
| | Sloanea zuliaensis | 2.10 |
| | Myrsine coriacea | 1.98 |

| cluster | species | ρij |
|---|---|---|
| C6 | Otoba novogranatensis | 13.07 |
| | Poulsenia armata | 10.91 |
| | Brosimum costaricanum | 9.24 |
| | Mortoniodendron anisophyllum | 9.06 |
| | Tetrathylacium macrophyllum | 8.07 |
| | Sorocea pubivena | 7.74 |
| | Meliosma allenii | 7.67 |
| | Calatola costaricensis | 7.20 |
| | Batocarpus costaricensis | 6.73 |
| | Vochysia gentryi | 6.71 |
| | Virola nobilis | 6.52 |
| | Carapa nicaraguensis | 6.13 |
| | Caryocar costaricense | 5.98 |
| | Brosimum utile | 5.27 |
| | Dendropanax caucanus | 5.16 |
| | Cleidion castaneifolium | 5.05 |
| | Elaeagia myriantha | 4.99 |
| | Alchornea costaricensis | 4.80 |
| | Stephanopodium costaricense | 4.75 |
| | Virola koschnyi | 4.71 |
| | Phyllanthus skutchii | 4.58 |
| | Miconia donaeana | 4.41 |
| | Zygia confusa | 4.22 |
| | Pleuranthodendron lindenii | 4.07 |
| | Andira inermis | 4.02 |
| | Chimarrhis parviflora | 3.93 |
| | Perebea hispidula | 3.65 |
| | Hirtella triandra | 3.60 |
| | Myriocarpa longipes | 3.60 |
| | Grias cauliflora | 3.27 |
| | Persea americana | 3.25 |
| | Chrysochlamys glauca | 3.17 |
| | Symphonia globulifera | 3.16 |
| | Apeiba tibourbou | 3.06 |
| | Pterocarpus rohrii | 3.05 |
| | Vitex cooperi | 2.84 |
| | Vochysia guatemalensis | 2.72 |
| | Talisia nervosa | 2.70 |
| | Swartzia simplex | 2.66 |
| | Quararibea asterolepis | 2.63 |
| | Ormosia subsimplex | 2.61 |
| | Pachira aquatica | 2.57 |
| | Ficus tonduzii | 2.51 |
| | Talisia macrophylla | 2.51 |
| | Zanthoxylum ekmanii | 2.46 |
| | Trichospermum galeottii | 2.44 |
| | Drypetes brownii | 2.32 |
| | Unonopsis pittieri | 2.29 |
| | Ficus costaricana | 2.21 |
| | Stenostomum turrialbanum | 2.21 |
| | Garcinia madruno | 2.18 |
| | Cecropia insignis | 2.13 |
| | Trichilia septentrionalis | 2.06 |
| | Inga oerstediana | 2.05 |
| | Clarisia biflora | 2.04 |

| cluster | species | pij |
| --- | --- | --- |
| C7 | Saurauia montana | 8.81 |
| | Inga punctata | 8.23 |
| | Viburnum costaricanum | 6.98 |
| | Heliocarpus appendiculatus | 6.71 |
| | Erythrina poeppigiana | 5.91 |
| | Beilschmiedia pendula | 5.17 |
| | Cecropia obtusifolia | 4.79 |
| | Aiouea montana | 4.69 |
| | Oreopanax xalapensis | 4.59 |
| | Alnus acuminata | 4.51 |
| | Croton skutchii | 4.51 |
| | Clethra lanata | 4.51 |
| | Cornutia pyramidata | 4.42 |
| | Tetrathylacium johansenii | 4.38 |
| | Quercus corrugata | 3.94 |
| | Nectandra reticulata | 3.78 |
| | Tabebuia rosea | 3.63 |
| | Nectandra membranacea | 3.40 |
| | Heliocarpus americanus | 3.29 |
| | Ficus maxima | 3.19 |
| | Myrcia splendens | 3.09 |
| | Myrsine coriacea | 2.76 |
| | Hura crepitans | 2.42 |
| | Persea caerulea | 2.18 |
| | Cedrela tonduzii | 2.15 |
| | Persea schiedeana | 2.12 |
| | Miconia donaeana | 2.11 |
| | Croton draco | 2.06 |

| cluster | species | pij |
| --- | --- | --- |
| C8 | Aiouea pittieri | 9.73 |
| | Sorocea trophoides | 9.58 |
| | Pseudolmedia glabrata | 9.36 |
| | Tapirira mexicana | 9.02 |
| | Clarisia racemosa | 8.94 |
| | Ardisia compressa | 8.81 |
| | Clarisia biflora | 7.93 |
| | Trema domingense | 7.89 |
| | Zinowiewia integerrima | 7.75 |
| | Laplacea fructicosa | 7.69 |
| | Licaria nitida | 7.13 |
| | Cojoba membranacea | 7.13 |
| | Podocarpus costaricensis | 7.13 |
| | Damburneya smithii | 6.86 |
| | Chomelia microloba | 6.73 |
| | Gymnanthes riparia | 6.68 |
| | Ficus obtusifolia | 6.61 |
| | Chionanthus panamensis | 6.31 |
| | Cordia megalantha | 6.12 |
| | Myrcianthes fragrans | 5.89 |
| | Ficus insipida | 5.80 |
| | Ficus crocata | 5.73 |
| | Amyris pinnata | 5.64 |
| | Ocotea stenoneura | 5.63 |
| | Lafoensia punicifolia | 4.80 |
| | Sorocea affinis | 4.73 |
| | Beilschmiedia pendula | 4.66 |
| | Hauya elegans | 4.64 |
| | Jacaratia dolichaula | 4.32 |
| | Cupania guatemalensis | 4.14 |
| | Zanthoxylum acuminatum | 4.02 |
| | Beilschmiedia costaricensis | 3.96 |
| | Terminalia oblonga | 3.90 |
| | Oreopanax xalapensis | 3.85 |
| | Sideroxylon portoricense | 3.31 |
| | Meliosma idiopoda | 2.79 |
| | Diospyros juruensis | 2.64 |
| | Heliocarpus appendiculatus | 2.35 |
| | Persea americana | 2.27 |
| | Heisteria concinna | 2.27 |
| | Lonchocarpus acuminatus | 2.20 |
| | Astronium graveolens | 2.20 |
| | Chrysophyllum brenesii | 2.10 |
| | Swartzia simplex | 2.05 |

| cluster | species | pij |
| --- | --- | --- |
| C9 | Psychotria berteroana | 8.70 |
| | Guarea chiricana | 7.60 |
| | Oreomunnea pterocarpa | 6.87 |
| | Ticodendron incognitum | 6.45 |
| | Calophyllum brasiliense | 5.70 |
| | Ocotea pullifolia | 5.32 |
| | Myrcia chytraculia | 5.17 |
| | Virola montana | 4.23 |
| | Perrottetia multiflora | 4.16 |
| | Billia rosea | 3.46 |
| | Guarea guidonia | 3.35 |
| | Xylopia sericea | 3.17 |
| | Sorocea pubivena | 3.01 |
| | Macrohasseltia macroterantha | 2.75 |
| | Cynometra retusa | 2.43 |
| | Quercus corrugata | 2.38 |
| | Brosimum lactescens | 2.36 |
| | Casearia sylvestris | 2.34 |
| | Inga pezizifera | 2.19 |
| | Unonopsis pittieri | 1.99 |

| cluster | species | pij |
|---|---|---|
| C10 | Mortoniodendron apetalum | 16.27 |
| | Cupania juglandifolia | 13.16 |
| | Diospyros juruensis | 12.68 |
| | Perrottetia longistylis | 12.45 |
| | Chromolucuma congestifolia | 12.40 |
| | Trophis racemosa | 12.21 |
| | Guarea kegelii | 11.74 |
| | Inga mortoniana | 10.90 |
| | Pachira aquatica | 10.61 |
| | Pentagonia costaricensis | 10.60 |
| | Sapium glandulosum | 10.30 |
| | Inga sapindoides | 10.14 |
| | Pseudolmedia mollis | 10.00 |
| | Tabernaemontana longipes | 9.95 |
| | Meliosma glabrata | 9.71 |
| | Ocotea dentata | 9.47 |
| | Zanthoxylum acuminatum | 8.50 |
| | Sapium rigidifolium | 8.37 |
| | Ficus hartwegii | 8.35 |
| | Cedrela tonduzii | 8.32 |
| | Citharexylum costaricense | 7.93 |
| | Warszewiczia uxpanapensis | 7.80 |
| | Lonchocarpus monteviridis | 7.63 |
| | Alchornea glandulosa | 7.59 |
| | Lunania mexicana | 7.53 |
| | Inga barbourii | 7.51 |
| | Ocotea insularis | 7.41 |
| | Chimarrhis parviflora | 6.99 |
| | Ficus maxima | 6.76 |
| | Ruagea glabra | 6.22 |
| | Meliosma idiopoda | 5.83 |
| | Inga marginata | 5.18 |
| | Inga oerstediana | 4.78 |
| | Nectandra reticulata | 4.09 |
| | Staphylea occidentalis | 3.90 |
| | Ocotea laetevirens | 3.68 |
| | Sorocea trophoides | 3.59 |
| | Hasseltia guatemalensis | 3.49 |
| | Drypetes brownii | 3.31 |
| | Cecropia insignis | 2.89 |
| | Trichilia martiana | 2.46 |
| | Alchornea latifolia | 2.35 |
| | Inga leonis | 2.10 |
| | Calatola costaricensis | 2.07 |

**Table S3.** The matrix of mean fractions of contribution to the cluster j from contributive species ($\rho_{ij} \geq 1.96$) that also contribute significantly to cluster $j'$ ($\lambda_{jj'}$). The grey values in the matrix represent $\lambda_{jj}$, the specificity of cluster $j$. These fractions are expressed as percentages with a vector $\lambda_j$ for a given cluster that sums to 1.

| Matrix λ jj' | | | | | | | | | | |
|---|---|---|---|---|---|---|---|---|---|---|
| | **C2** | **C9** | **C3** | **C7** | **C1** | **C5** | **C6** | **C8** | **C4** | **C10** |
| **C2** | 0.88 | 0.02 | 0.00 | 0.00 | 0.07 | 0.01 | 0.02 | 0.00 | 0.00 | 0.01 |
| **C9** | 0.08 | 0.69 | 0.00 | 0.03 | 0.09 | 0.01 | 0.05 | 0.00 | 0.05 | 0.00 |
| **C3** | 0.00 | 0.00 | 0.99 | 0.00 | 0.00 | 0.00 | 0.00 | 0.00 | 0.00 | 0.00 |
| **C7** | 0.00 | 0.03 | 0.00 | 0.78 | 0.00 | 0.03 | 0.02 | 0.07 | 0.02 | 0.05 |
| **C1** | 0.04 | 0.02 | 0.00 | 0.00 | 0.88 | 0.00 | 0.06 | 0.00 | 0.00 | 0.01 |
| **C5** | 0.00 | 0.01 | 0.00 | 0.01 | 0.00 | 0.85 | 0.02 | 0.03 | 0.03 | 0.06 |
| **C6** | 0.02 | 0.01 | 0.00 | 0.00 | 0.10 | 0.02 | 0.80 | 0.01 | 0.00 | 0.03 |
| **C8** | 0.00 | 0.00 | 0.01 | 0.04 | 0.01 | 0.03 | 0.02 | 0.86 | 0.00 | 0.04 |
| **C4** | 0.00 | 0.02 | 0.00 | 0.01 | 0.00 | 0.06 | 0.01 | 0.00 | 0.90 | 0.01 |
| **C10** | 0.01 | 0.00 | 0.01 | 0.01 | 0.02 | 0.08 | 0.03 | 0.02 | 0.00 | 0.82 |

**Table S4.** List of contributive species that contribute to more than one in situ cluster, i.e. with a relative contribution $\hat{\rho}^{+}_{ij}$ of a species $i$ to a cluster $j$ less than 1 : 96 species contribute to more than one cluster, with 7 contributing to three clusters and 89 to two clusters, based on $\rho_{ij} \geq 1.96$.

| **Species** | **Clusters** | | |
|---|---|---|---|
| *Cecropia insignis* | C5 | C6 | C10 |
| *Brosimum lactescens* | C2 | C9 | C1 |
| *Sorocea pubivena* | C9 | C1 | C6 |
| *Pterocarpus rohrii* | C2 | C5 | C6 |
| *Virola koschnyi* | C2 | C1 | C6 |
| *Unonopsis pittieri* | C2 | C9 | C6 |
| *Swartzia simplex* | C1 | C6 | C8 |
| *Virola sebifera* | C2 | C1 | |
| *Simarouba amara* | C2 | C1 | |
| *Guarea guidonia* | C9 | C5 | |
| *Vochysia guatemalensis* | C1 | C6 | |
| *Dendropanax arboreus* | C2 | C1 | |
| *Ficus maxima* | C7 | C10 | |
| *Ocotea pullifolia* | C2 | C9 | |
| *Tapirira guianensis* | C2 | C1 | |
| *Staphylea occidentalis* | C5 | C10 | |
| *Inga pezizifera* | C2 | C9 | |
| *Calophyllum brasiliense* | C9 | C1 | |
| *Inga oerstediana* | C5 | C10 | |
| *Nectandra reticulata* | C7 | C10 | |
| *Astronium graveolens* | C3 | C8 | |
| *Apeiba tibourbou* | C1 | C6 | |
| *Heliocarpus appendiculatus* | C7 | C8 | |
| *Dussia macroprophyllata* | C2 | C1 | |
| *Drypetes brownii* | C6 | C10 | |
| *Pouteria torta* | C2 | C1 | |
| *Xylopia sericophylla* | C2 | C1 | |
| *Cupania glabra* | C2 | C1 | |
| *Brosimum guianense* | C2 | C1 | |
| *Inga thibaudiana* | C2 | C1 | |
| *Alchornea latifolia* | C5 | C10 | |
| *Beilschmiedia costaricensis* | C5 | C8 | |
| *Maranthes panamensis* | C2 | C1 | |
| *Pachira aquatica* | C6 | C10 | |
| *Ocotea laetevirens* | C2 | C10 | |
| *Myrcia chytraculia* | C9 | C5 | |
| *Persea americana* | C6 | C8 | |
| *Hirtella triandra* | C2 | C6 | |
| *Garcinia madruno* | C1 | C6 | |
| *Brosimum utile* | C1 | C6 | |
| *Xylopia sericea* | C9 | C1 | |

| | | |
|---|---|---|
| *Symphonia globulifera* | C1 | C6 |
| *Caryocar costaricense* | C1 | C6 |
| *Miconia brenesii* | C5 | C4 |
| *Calatola costaricensis* | C6 | C10 |
| *Guatteria oliviformis* | C5 | C4 |
| *Trichilia martiana* | C3 | C10 |
| *Zanthoxylum acuminatum* | C8 | C10 |
| *Inga barbourii* | C5 | C10 |
| *Warszewiczia uxpanapensis* | C5 | C10 |
| *Myrsine coriacea* | C7 | C4 |
| *Cordia cymosa* | C2 | C5 |
| *Grias cauliflora* | C2 | C6 |
| *Anaxagorea crassipetala* | C2 | C1 |
| *Chimarrhis parviflora* | C6 | C10 |
| *Billia rosea* | C9 | C4 |
| *Lonchocarpus monteviridis* | C5 | C10 |
| *Quararibea asterolepis* | C3 | C6 |
| *Meliosma idiopoda* | C8 | C10 |
| *Hasseltia guatemalensis* | C5 | C10 |
| *Beilschmiedia pendula* | C7 | C8 |
| *Oreopanax xalapensis* | C7 | C8 |
| *Chrysophyllum brenesii* | C1 | C8 |
| *Sorocea trophoides* | C8 | C10 |
| *Ruagea glabra* | C5 | C10 |
| *Sapium rigidifolium* | C5 | C10 |
| *Heliocarpus americanus* | C7 | C5 |
| *Pleurothyrium palmanum* | C5 | C4 |
| *Terminalia oblonga* | C5 | C8 |
| *Damburneya cufodontisii* | C9 | C4 |
| *Miconia durandii* | C5 | C4 |
| *Vitex cooperi* | C2 | C6 |
| *Vochysia gentryi* | C1 | C6 |
| *Cedrela tonduzii* | C7 | C10 |
| *Miconia donaeana* | C7 | C6 |
| *Persea schiedeana* | C9 | C7 |
| *Perrottetia multiflora* | C9 | C6 |
| *Quercus corrugata* | C9 | C7 |
| *Viburnum costaricanum* | C7 | C4 |
| *Carapa nicaraguensis* | C1 | C6 |
| *Tetrathylacium macrophyllum* | C1 | C6 |
| *Virola montana* | C9 | C1 |
| *Diospyros juruensis* | C8 | C10 |
| *Ocotea stenoneura* | C5 | C8 |
| *Sideroxylon portoricense* | C5 | C8 |
| *Alchornea glandulosa* | C5 | C10 |
| *Perebea hispidula* | C1 | C6 |
| *Trichospermum galeottii* | C1 | C6 |
| *Dendropanax caucanus* | C6 | C4 |
| *Macrohasseltia macroterantha* | C9 | C1 |
| *Croton draco* | C7 | C5 |
| *Virola nobilis* | C1 | C6 |
| *Persea caerulea* | C7 | C5 |
| *Talisia macrophylla* | C5 | C6 |
| *Inga leonis* | C5 | C10 |
| *Ilex lamprophylla* | C5 | C4 |

**Table S5.** Mean values and standard deviation (SD) of the environmental variables used to characterise the 7 modelled forest ecosystems. q25, q50 and q75 represent the 25th, 50th, and 75th percentiles of the variables measured across the forest segments, respectively. Topographic variables: DEM, representing elevation in meters, and slope, expressed in degrees. Climatic variables: PRSea, the coefficient of variation of precipitation seasonality, and anPR, the mean annual precipitation, expressed in mm/year. Edaphic variables: pH and Cation Exchange Capacity (CEC) expressed in mmolc/kg, at a depth of 30 cm. Vegetation dynamics variable: NDWIw, Normalised Difference Water Index of wet season. Importance values are derived from the permutation of variables in the model.

| Clusters | q50.DEM | | Q75.anPR | | Q25.PRSea | | q25.NDWIw | | q50.pH30 | | q25.Slope | | q50.CEC30 | |
|---|---|---|---|---|---|---|---|---|---|---|---|---|---|---|
| | mean | sd | mean | sd | mean | sd | mean | sd | mean | sd | mean | sd | mean | sd |
| C1 | 504 | 361 | 3611 | 707 | 51 | 13 | 0.31 | 0.11 | 5.21 | 0.62 | 10 | 8 | 134 | 44 |
| C2 | 141 | 205 | 3522 | 726 | 34 | 9 | 0.31 | 0.07 | 5.27 | 0.62 | 4 | 4 | 132 | 49 |
| C3 | 258 | 261 | 2308 | 542 | 75 | 14 | 0.22 | 0.09 | 5.73 | 0.69 | 8 | 7 | 212 | 60 |
| C4 | 2269 | 463 | 3804 | 614 | 45 | 10 | 0.37 | 0.08 | 5.12 | 0.17 | 20 | 8 | 192 | 39 |
| C5 | 1142 | 509 | 3616 | 547 | 48 | 13 | 0.37 | 0.05 | 5.34 | 0.2 | 14 | 7 | 171 | 45 |
| C6 | 296 | 206 | 3666 | 667 | 65 | 8 | 0.39 | 0.03 | 5.28 | 0.29 | 13 | 8 | 183 | 42 |
| C7 | 1360 | 372 | 3245 | 549 | 55 | 12 | 0.26 | 0.08 | 5.29 | 0.31 | 16 | 8 | 162 | 33 |

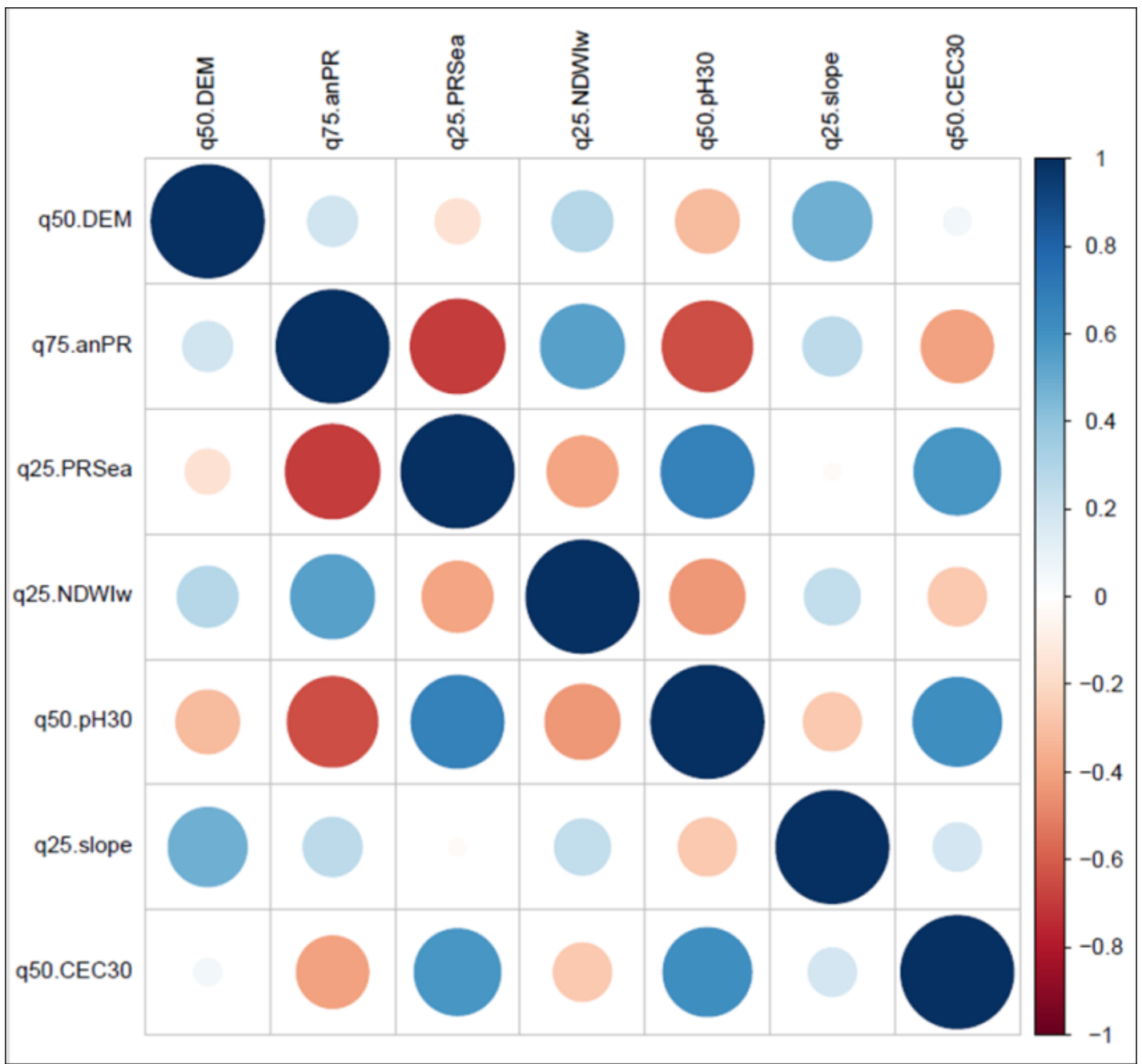

**Figure S3**. Correlation matrix of key variables selected for the Random Forest model from Pearson correlation: Topographic variables. q25, q50 and q75 represent the 25th, 50th, and 75th percentiles of the variables measured across the forest segments, respectively. DEM, representing elevation in meters, and Slope, expressed in degrees. Climatic variables: PRSea, the coefficient of variation of precipitation seasonality, and anPR, the mean annual precipitation, expressed in mm/year. Edaphic variables: pH and CEC, Cation Exchange Capacity expressed in mmolc/kg, at a depth of 30 cm. Vegetation dynamics variable : NDWIw, Normalized Difference Water Index of wet season.

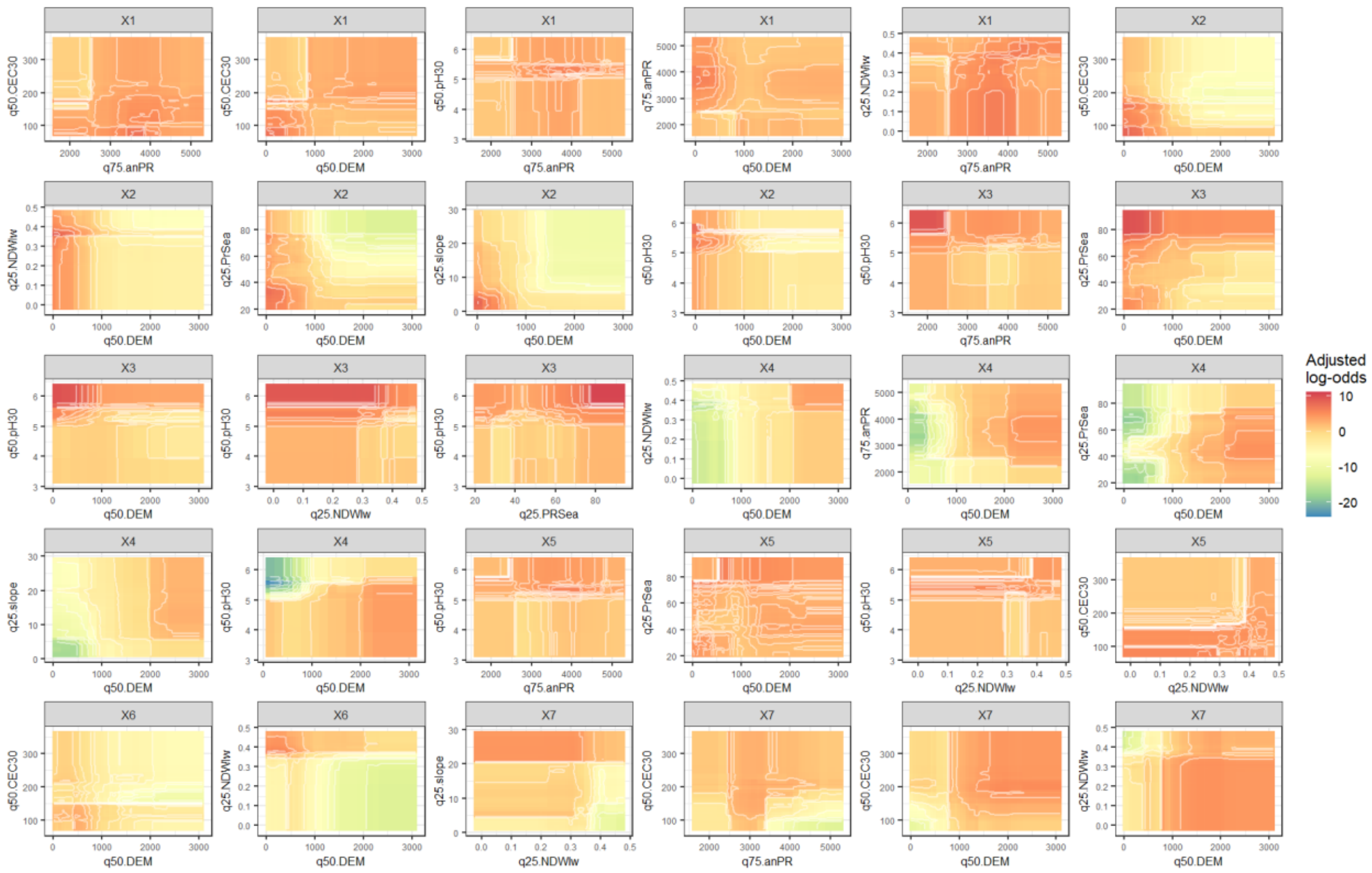

**Figure S4.** Partial dependence plots between two variables to analyse the main marginal effects of the most important predictive variables for each cluster in adjusted log-odds. Positive adjusted log-odds indicate a high probability of classification into the given cluster, zero represents a 50% probability, and negative values indicate a low or near zero probability of classification.

**Table S6.** Distribution of plots by number and proportion (%) along the altitudinal gradient in Costa Rica.

| Plots | | Altitudinal range |
|---|---|---|
| **Number** | **Proportion** | |
| 250 | 69.64 | [0, 500) |
| 51 | 14.21 | [500, 1000) |
| 14 | 3.90 | [1000, 1500) |
| 42 | 11.70 | [1500, 2000) |
| 2 | 0.56 | [2000, 2500) |

**Table S7**. Correspondence table between local and national forest characterisations of the lowland forests of Osa Peninsula. The local characterisation, conducted by Hofhansl et al. (2019) from some sites included in our analysis, highlights the key role of tree species that contribute to the national forest ecosystems present in this region. In these sites, ecosystem WSE dominates, although ecosystems LWE-C and TWPE-P are also represented. They identify four forest-types (Ridge, Salope, Ravine and Secondary) sharing many dominant species based on the 10 most dominant species of each type. This table illustrates the complexity of delineating forest-types in the peninsula, reflected by the overlap of contributive species shared by different forest ecosystems identified at the national level.

| Dominant Species | Local characterisation | National Characterisation |
|---|---|---|
| *Vochysia ferruginea* | Ridge forests | WSE |
| *Pourouma bicolor* | Ridge forests | WSE, LWE-C |
| *Compsoneura excelsa* | Ridge and Slope forests | WSE, LWE-C, TWPE-P |
| *Mabea occidentalis* | Ridge and, Slope forests | WSE, LWE-C, TWPE-P |
| *Tapirira guianensis* | Ridge and Slope forests | WSE, LWE-C, TWPE-P |
| *Otoba novogranatensis* | Slope and Ravine forests | TWPE-P |
| *Sorocea pubivena* | Slope and Ravine forests | WSE, TWPE-P |
| *Goethalsia meiantha* | Ravine and Secondary forests | LWE-C |
| *Apeiba tibourbou* | Secondary forests | WSE, TWPE-P |
| *Hieronyma alchorneoides* | Secondary forests | WSE, TWPE-P |
| *Castilla tunu* | Secondary forests | WSE, TWPE-P |
| *Alchornea costaricensis* | Secondary forests | WSE, TWPE-P |
| *Tetrathylacium macrophyllum* | Slope, Ravine and Secondary forests | WSE, TWPE-P |
| *Symphonia globulifera* | Ridge, Slope and Ravine forests | WSE, TWPE-P |
| *Carapa nicaraguensis* | Ridge, Slope, Ravine and secondary forests | WSE, LWE-C, TWPE-P |

**Table S8**. Description based on botanical expertise of the seven main modelled national forest ecosystems.

| Ecosystems | Description |
|---|---|
| Wet Seasonal Evergreen forest (WSE) | This forest ecosystem is characterised by a high composition of tree species, typical of old-growth forests in a well-preserved state. Its indicator species commonly occupy the canopy or emerge above it and are significantly abundant within the ecosystem. While no particular species dominates, species diversity and abundance are heterogeneous. This species composition and/or mix is consistent with the altitudinal range in which they are distributed, mainly between 0 and 700 m, and they predominantly occupy very humid seasonal lowland forests. Biogeographically, this ecosystem has two cores of tree diversity, the most diverse located in the Osa Peninsula and another, more isolated, restricted to the northernmost region of Costa Rica. Both cores (northern zone/Osa Peninsula) share a significant number of tree species, yet also contain unique, non-overlapping species specific to each biogeographical region. These ecosystems typically occupy areas with rather irregular topography. |
| Lowland Wet Evergreen forest of Caribbean slope (LWE-C) | This forest ecosystem is characterised by a tree species composition that is clearly dominated by the abundance or presence of Pentaclethra macroloba. Forests with a high presence of Pentaclethra macroloba exhibit a very distinctive associated tree diversity, both in the canopy and the understory, with a strong dominance of arboreal or shrubby palms. Although this ecosystem has high species diversity, other tree species (besides Pentaclethra macroloba) can also become dominant in the overall forest structure. The species composition of this ecosystem results from forests subject to intervention or logging, as evidenced by the presence of a number of fast-growing species interspersed, which are more typical of disturbed forests or late-secondary to old secondary forests. This ecosystem prefers areas with relatively flat or undulating topography, as well as alluvial plains. |
| Lowland Dry-to-Moist Deciduous-to-Semi-deciduous forest (LDM-DS) | This forest ecosystem is characterised by a composition of tree species that are typical of, or commonly found in, deciduous or semi-deciduous dry forests, interspersed with some evergreen species scattered throughout the landscape or associated with riparian forests. Its diversity is heterogeneous but not particularly high, as it exhibits strong dominance by species from the Fabaceae family (primarily), as well as Bignoniaceae, Malvaceae, Burseraceae, Anacardiaceae, among others. This ecosystem experiences the highest levels of climatic seasonality in the country, with the greatest number of dry months. As a result, this climatic factor is the primary driver of the dominant floristic pattern. Additionally, this ecosystem has undergone a long history of anthropogenic use and impacts (particularly fire), which have significantly altered its original species composition. In general, this ecosystem hosts a tree species composition with a broad geographical distribution along the Pacific coast of Mesoamerica. In Costa Rica, in particular, it is closely associated with the climatic seasonality gradient along the Pacific coast. Notably, some species from this ecosystem also occur in other, distant and ecologically distinct ecosystems (secondary forests) within the country, largely as a consequence of the historical movement of cattle ranching across the landscape. |
| Mountain Oak Rainforest (MOR) | This forest ecosystem is characterised by a tree species composition largely dominated by oak species (Quercus spp.), forming the well-known associations called "oak forests", where, depending on the altitudinal gradient, one, two, or more Quercus species dominate the forest structure. The sampling plots were mainly concentrated within the 1500–2000 m elevation band, primarily on the Pacific slope, where the forests are seasonal evergreen. Consequently, the recorded composition is representative of this altitudinal range. Additionally, most of the sampled plots correspond to old-growth forests. The largest expanse of oak forests typically occurs above 2000 m. |
| Premontane-to-mountain Mixed-to-evergreen Cloud forest of Caribbean slope (PMC-C) | This forest ecosystem is characterised by a composition of tree species from mid-elevations (1410–1957 m, based on the sampling) on the Caribbean slope. The inclusion of several plots below 500 m, all classified as Secondary Forest, affects the definition and characterisation of this ecosystem's tree composition. Moreover, most of these Secondary Forest plots actually belong to the Caribbean lowland ecosystem. |
| Transitional wet premontane evergreen forest of the Pacific slope (TWPE-P) | *Otoba novogranatensis* and *Poulsenia armata*, indicators of forest maturity, are the most contributive species. However, this ecosystem is found in forests that have been disturbed by selective logging, which explains the presence of many species typical of secondary or disturbed forests among the contributive species. This ecosystem, primarily located in the Osa Peninsula, represents an extension of the LWE-C ecosystem, with many common contributive species. It is likely that this discrimination is due to the inclusion in the analysis of 11 LWE-C plots located above 700 m in altitude, 8 of which come from secondary forests. The tree composition model above this altitude differs significantly from the LWE-C type. |
| Premontane-to-mountain Mixed-to-evergreen Cloud forest of Pacific slope (PMC-P) | This forest ecosystem is characterised by a tree species composition dominated by secondary forest species, as the majority or nearly all of the plots belong to SF. Additionally, most plots are located in the mid-elevation band (1111 m) on the Pacific slope, where the vegetation cover is highly fragmented due to coffee cultivation. This elevation band also experiences a climate with a strongly marked seasonality. |